\documentclass[aps,prl,reprint,superscriptaddress,  onecolumn]{revtex4-2}
\usepackage{gensymb}
\usepackage{amsmath}
\usepackage{nccmath}
\usepackage{graphics}
\usepackage{epsfig}
\usepackage{bm}
\usepackage{dcolumn}
\usepackage{float}
\usepackage{nicefrac}
\usepackage{siunitx}
\usepackage[colorlinks=true, hyperindex, breaklinks, linkcolor=blue, urlcolor=blue, citecolor=blue]{hyperref} 
\renewcommand{\vec}[1]{\mathbf{#1}}

\begin{document}

\title{Nanoscale Electric Field Imaging with an Ambient Scanning Quantum Sensor Microscope}

\author{Ziwei Qiu}
\email[ziweiqiu29@gmail.com]{}
\affiliation{Department of Physics, Harvard University, Cambridge, MA 02138, USA}
\affiliation{John A. Paulson School of Engineering and Applied Sciences, Harvard University, Cambridge, MA 02138, USA}
\author{Assaf Hamo}
\email[Present address: Department of Physics, Bar-Ilan University, Ramat Gan, Israel]{}
\affiliation{Department of Physics, Harvard University, Cambridge, MA 02138, USA}
\author{Uri Vool}
\email[Present address: Max Planck Institute for Chemical Physics of Solids, Dresden, Germany]{}
\affiliation{Department of Physics, Harvard University, Cambridge, MA 02138, USA}
\affiliation{John Harvard Distinguished Science Fellows Program, Harvard University, Cambridge MA 02138, USA}
\author{Tony X. Zhou}
\email[Present address: Research Laboratory of Electronics, Massachusetts Institute of Technology, Cambridge, MA, USA]{}
\affiliation{Department of Physics, Harvard University, Cambridge, MA 02138, USA}
\affiliation{John A. Paulson School of Engineering and Applied Sciences, Harvard University, Cambridge, MA 02138, USA}
\author{Amir Yacoby}
\email[Correspondence to: yacoby@physics.harvard.edu]{}
\affiliation{Department of Physics, Harvard University, Cambridge, MA 02138, USA}
\affiliation{John A. Paulson School of Engineering and Applied Sciences, Harvard University, Cambridge, MA 02138, USA}

\date{\today}

\begin{abstract}
Nitrogen-vacancy (NV) center in diamond is a promising quantum sensor with remarkably versatile sensing capabilities. While scanning NV magnetometry is well-established, NV electrometry has been so far limited to bulk diamonds. Here we demonstrate imaging external alternating (AC) and direct (DC) electric fields with a single NV at the apex of a diamond scanning tip under ambient conditions. A strong electric field screening effect is observed at low frequencies due to charge noise on the surface. We quantitatively measure its frequency dependence, and overcome this screening by mechanically oscillating the tip for imaging DC fields. Our scanning NV electrometry achieved an AC E-field sensitivity of \SI{26}{\,mV/\mu m/\sqrt{Hz}},  a DC E-field gradient sensitivity of \SI{2}{\,V/\mu m^2/\sqrt{Hz}}, and sub-\SI{100}{\,nm} resolution limited by the NV-sample distance. Our work represents an important step toward building a scanning-probe-based multimodal quantum sensing platform. 
\end{abstract}

\maketitle

\section*{Introduction}
The recent decade has witnessed exciting developments in quantum science and technology, where controlled quantum systems are employed to carry out tasks that are challenging for existing classical techniques. Quantum sensing is a rapidly-growing branch \cite{degen2017quantum}, which exploits the fragility of quantum states to detect small external signals with high sensitivity. It has many potential applications in the real world, such as geophysical navigation \cite{little2021passively, soshenko2021nuclear}, disease diagnostics \cite{miller2020spin, li2021sars} and discovery of new materials \cite{thiel2019probing,ku2020imaging, vool2021imaging}. The nitrogen-vacancy (NV) center in diamond is one of the most promising quantum sensors \cite{schirhagl2014nitrogen}. Its electron spin has a long coherence time even under ambient conditions. It is sensitive to a variety of signals, such as magnetic fields \cite{maze2008nanoscale, mamin2013nanoscale, hong2013nanoscale, casola2018probing}, electric fields \cite{dolde2011electric, michl2019robust, yang2020vector, li2020nanoscale, block2021optically, barson2021nanoscale, bian2021nanoscale}, temperature \cite{kucsko2013nanometre, neumann2013high, toyli2013fluorescence, choi2020probing} and pressure \cite{doherty2014electronic, ivady2014pressure, kehayias2019imaging}. Integrating this atomic-sized versatile quantum sensor into a scanning probe microscope further allows mapping external signals with nanoscale resolution \cite{maletinsky2012robust, rondin2012nanoscale, pelliccione2016scanned}. This holds a great potential for probing condensed matter physics. In particular, imaging both magnetic and electric fields can provide a unique insight to strongly correlated matter \cite{wu2018hubbard} and multiferroic materials \cite{mundy2016atomically}. It also has interdisciplinary applications, such as imaging charge and spin phenomena in chemistry and biology \cite{schirhagl2014nitrogen, barry2016optical, liu2017scheme, petrini2020quantum, webb2021detection}. 

Scanning NV magnetometry, as a well-established technique \cite{maletinsky2012robust}, has been utilized to image magnetic materials \cite{tetienne2014nanoscale, tetienne2015nature, gross2017real, thiel2019probing}, hydrodynamic current flows \cite{ku2020imaging, vool2021imaging}, skyrmion structures \cite{dovzhenko2018magnetostatic} and vortices in superconductors \cite{thiel2016quantitative}. Recently, scanning NV electrometry has also been achieved, where fixed NVs in bulk diamonds are used to image electric fields from a conductive scanning tip \cite{barson2021nanoscale, bian2021nanoscale}. However, NV electrometry has not yet been demonstrated in a diamond scanning tip, which is desired for imaging an arbitrary sample. Major challenges arise from the qubit's weak coupling strength to electric fields, a relatively short coherence time in nanostructures compared to bulk diamonds, and electric field screening by the diamond surface \cite{oberg2020solution, stacey2019evidence, mertens2016patterned}. In this work, we demonstrate utilizing a shallow NV at the apex of a diamond nanopillar to image external AC and DC electric fields under ambient conditions. Dynamical decoupling sequences are used to extend the NV coherence times \cite{de2010universal}. We achieve an AC electric field sensitivity of \SI{26}{\,mV/\mu m/\sqrt{Hz}} and sub-\SI{100}{\,nm} spatial resolution limited by the NV-sample distance. A strong electric field screening effect is observed at low frequencies, likely caused by mobile charges on the diamond surface \cite{oberg2020solution}. We quantitatively characterize its frequency dependence, that reveals a resistive-capacitive (RC) time constant of our diamond tip surface $\sim$ \SI{\,30}{\,\mu s}. To overcome this screening effect, in order to image DC electric fields, we oscillate the diamond probe at a frequency of $\sim$ \SI{190}{\,kHz}, and synchronize the quantum sensing pulse sequences with the mechanical motion. Hence, a local spatial gradient is upconverted to an AC signal, allowing for $T_2$-limited DC sensing. This motion-enabled imaging technique has been explored in various scanning measurements \cite{hong2012coherent, kolkowitz2012coherent, teissier2014strain, halbertal2017imaging, wood2018t, huxter2022scanning}, and here we apply it to scanning NV electrometry. We achieve a DC electric field gradient sensitivity of \SI{2}{\,V/\mu m^2/\sqrt{Hz}}. Our results pave the way for building a scanning-probe-based multi-modal quantum sensing platform. 

\section*{Results}
\subsection*{Experimental Apparatus and NV Electrometry}
\begin{figure}[b]
\centering
\includegraphics[scale=0.55]{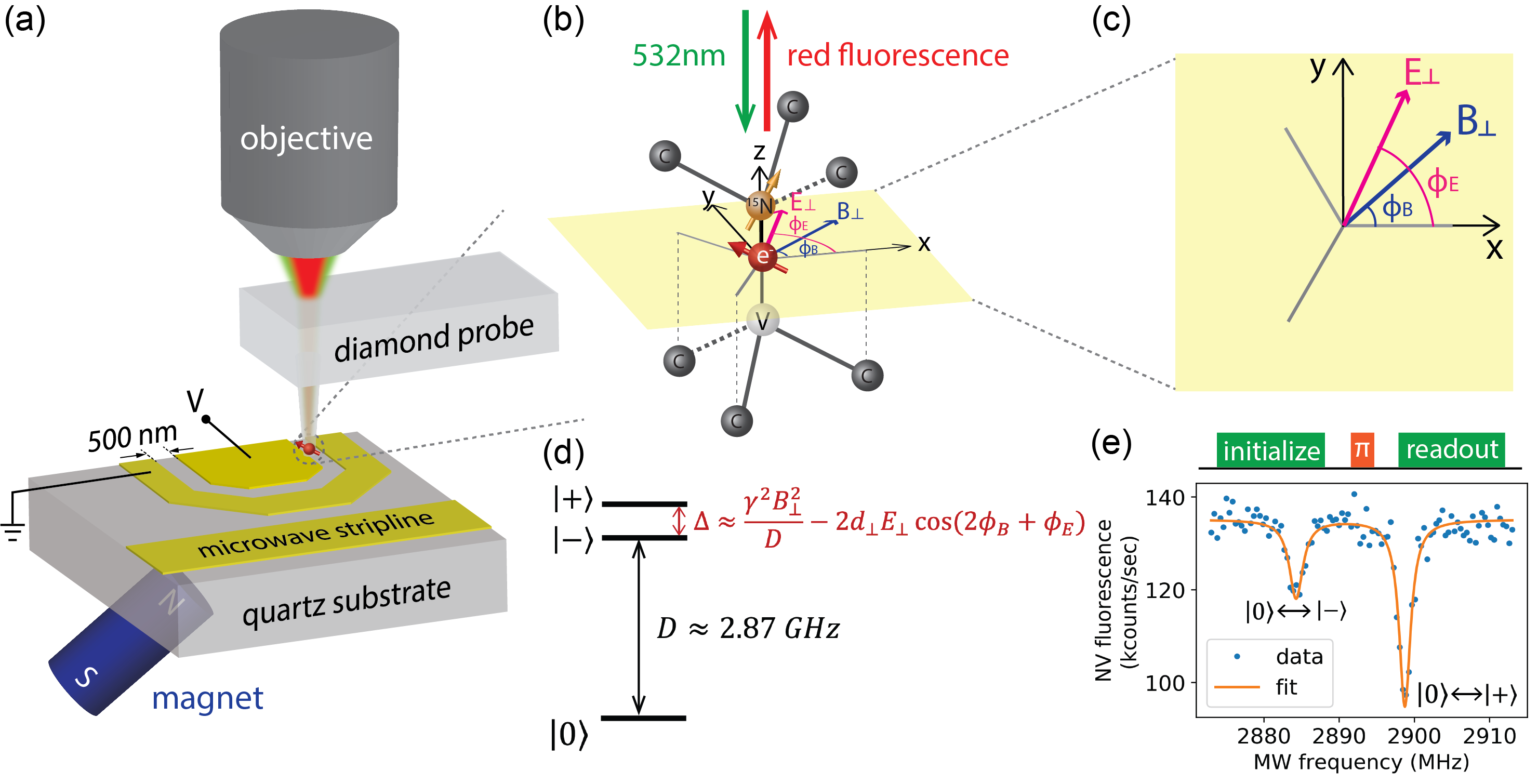} 
\caption{\label{fig:setup} Experimental setup and NV center. (a) Schematic of the experimental apparatus showing the confocal objective lens, a diamond probe, a U-shaped gold structure and microwave stripline fabricated on a quartz substrate, and an external magnet. The NV center is located at the apex of the diamond tip at a depth of $\sim$ 40 nm. The tip has a diameter of 300 nm and hovers above the sample. The NV-sample distance is typically $\lesssim100$ nm. The gap between the two electrodes is 500 nm, and Au is \SI[separate-uncertainty=true]{150\pm5}{nm} thick. (b) NV center in the presence of a perpendicular magnetic field, denoted by $\vec{B_\perp}$ in the XY plane in yellow. The $\hat{x}$ direction is defined by the projection of one of the carbon atoms. $E_{\perp}$ is the electric field component in the XY plane, which causes Stark shifts of the NV spin states. $\phi_B$ and $\phi_E$ denote the azimuth angle of $B_{\perp}$ and $E_{\perp}$ relative to the $\hat{x}$-axis. (c) Zoom-in of the XY plane. Grey lines represent the projections of carbon atoms. (d) NV electron spin energy level under $\vec{B_\perp}$. The $|\pm\rangle$ states are superpositions of $|m_S=\pm1\rangle$ (see main text) and sensitive to $E_{\perp}$. $d_\perp =$ \SI[separate-uncertainty=true]{0.17\pm 0.03}{\,MHz/V/\mu m} is the transverse electric field coupling constant and $\gamma$=2.8 MHz/G is the electron spin gyromagnetic ratio. (e) Top panel: optical and microwave sequence for pulsed optically detected magnetic resonance (ODMR) measurements. Bottom panel: Pulsed ODMR spectrum showing transitions between $|0\rangle$ and $|\pm \rangle$ under $B_\perp \approx$  73 G. The driving efficiencies depend on the MW field polarization. The $|0\rangle$-to-$|+\rangle$ transition is used throughout this work for electric field sensing.} 
\end{figure}

Our home-built diamond-NV scanning setup combines a confocal microscope and an atomic force microscope (AFM) operating in ambient conditions, as sketched in Fig.~\ref{fig:setup}a. The diamond probe and the sample can be independently scanned by piezoelectric nanopositioners. The probe is attached to a quartz crystal tuning fork, allowing for frequency modulation AFM (FM-AFM). The probe oscillates in a motion parallel to the sample surface, and the AFM operates in contact mode. To improve the scanning robustness, we make multiple nanopillars on a single diamond probe, and use different pillars for AFM contact and quantum sensing separately. The NV, at the apex of the sensing pillar, is kept at a fixed distance from the sample. The distance is controlled by the tilting angle of the probe. More details on AFM scanning control can be found in Supplementary Note 2E. To test the electrometry capabilities of our system, we fabricate a U-shaped gold structure on a quartz substrate and use the NV to map out its electric field field distribution. An on-chip microwave (MW) stripline delivers MW fields to manipulate the NV spin states. A permanent magnet from beneath exerts a bias magnetic field at the NV. 
 
Our electrometry measurements were performed under a magnetic field oriented perpendicular to the NV axis, denoted by $B_{\perp}$. The coordinate system is depicted in Fig.~\ref{fig:setup}b, where $\hat{z}$ is along the NV axis and projection of one carbon atom in the perpendicular plane defines the $\hat{x}$ axis. Under $B_{\perp}$, the NV electron spin eigenstates are $|0\rangle \approx |m_S = 0\rangle$, $|\pm \rangle \approx \frac{1}{\sqrt{2}}(|m_S=1\rangle \pm e^{2i\phi_B}|m_S=-1\rangle)$\cite{dolde2011electric, qiu2021nuclear}, where $\phi_B$ is the angle between $B_{\perp}$ and the $\hat{x}$ axis (see Fig.~\ref{fig:setup}c). Electric fields induce Stark shifts of the $|\pm\rangle$ energy levels. The splitting between the $|\pm\rangle$ states is approximately $\Delta \approx \frac{\gamma^2B_\perp^2}{D}-2d_\perp E_\perp \cos{(2\phi_B+\phi_E)}$, where $\phi_B$ and $\phi_E$ are the angles of magnetic and electric fields in the transverse plane as shown in Fig.~\ref{fig:setup}c, $D_{gs}\approx$ 2.87 GHz is the zero-field splitting (ZFS), $\gamma_B$= 2.8 MHz/G is the electron spin gyromagnetic ratio and $d_\perp =$ \SI[separate-uncertainty=true]{0.17\pm 0.03}{\,MHz/V/\mu m} is the transverse electric field coupling strength \cite{van1990electric}. We ignore the coupling to $E_z$ since the longitudinal coupling strength $d_\parallel=$ \SI[separate-uncertainty=true]{3.5\pm0.2}{\times10^{-3}\,MHz/V/\mu m} \cite{van1990electric} is much smaller. Strain is measured to be negligible in our diamond probe for the scope of this work.

We choose to work with $^{15}$NV under $B_\perp>70$ G  \cite{qiu2021nuclear}. The splitting between the $|\pm\rangle$ states is therefore much larger than the hyperfine coupling $\approx$ 3 MHz \cite{ohno2012engineering}, and both nuclear spin sublevels ($m_I=\pm1/2$) are sensitive to electric fields. Consequently, the full NV optical contrast contributes to the signal without the need of resolving hyperfine states. We use NVs with moderate $T_2^{*}$ dephasing times ($T_2^{*}\sim$1.5 $\mu s$, see Supplementary Note 2A) and apply relatively strong MW power to drive transitions between spin states ($\pi$ pulse duration $\sim$ 100 ns).  These are in contrast to NV electrometry performed under a weak magnetic field where typically $B_\perp<$ 20 G \cite{dolde2011electric, michl2019robust}. A detailed comparison between $^{15}$NV and $^{14}$NV electrometry under different magnetic field regimes can be found in Supplementary Note 1B. 

Fig.~\ref{fig:setup}d shows a pulsed optically detected magnetic resonance (ODMR) spectrum under $B_\perp \approx$ \SI[separate-uncertainty=true]{73}{G}. The transition efficiencies between $|0\rangle$ and $|\pm\rangle$ depend on the linear polarization of the MW fields (see Supplementary Note 1B). We use the $|0\rangle$-to-$|+\rangle$ transition throughout this work for electric field sensing.

\subsection*{Frequency Dependence of the Electric Field Screening} 

\begin{figure}[b]
\centering
\includegraphics[scale=0.61]{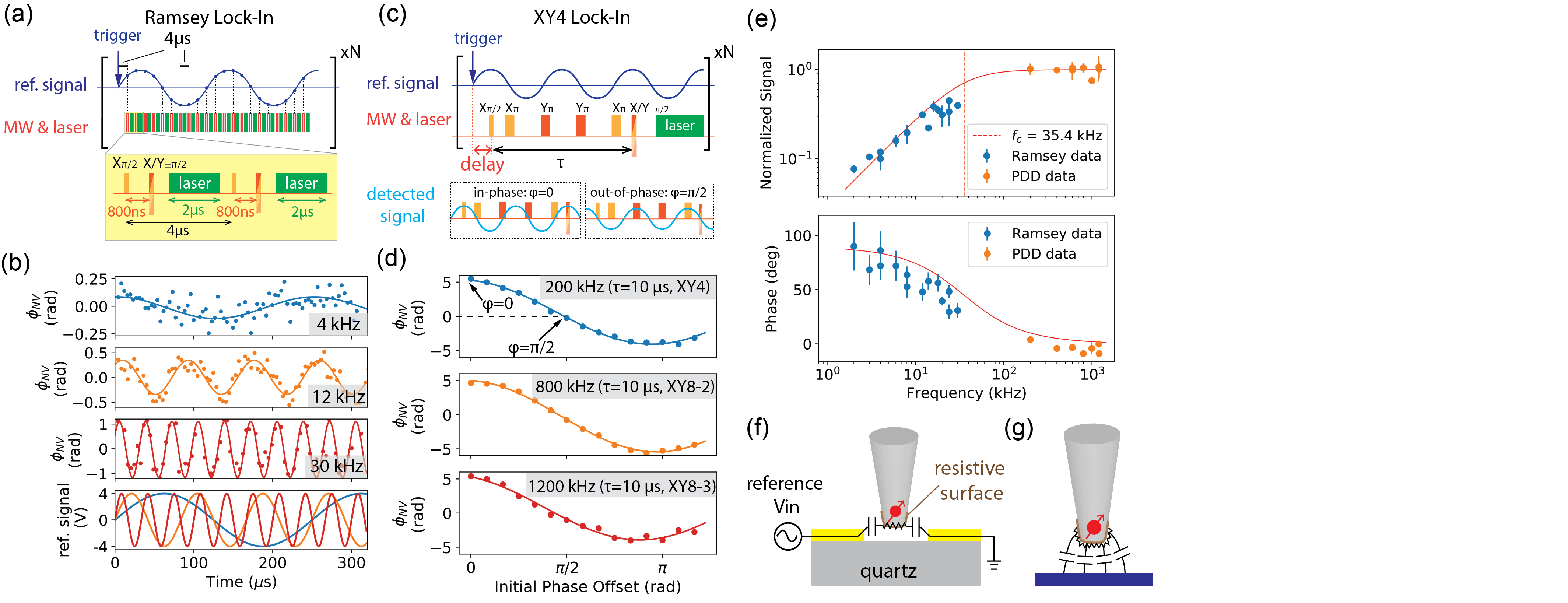}
\caption{\label{fig:screening} Frequency dependence of the electric field screening effect. (a) Top panel: Ramsey-based lock-in detection sequence used at low frequencies. A train of Ramsey measurements is synchronized with the external reference signal, and samples the local electric field every 4 $\mu s$. Bottom panel: Zoom-in of two Ramsey sequences. (b) Top three panels: Time traces of the Ramsey phase signal $\phi_{NV}$ at reference frequencies of 4 kHz, 12 kHz and 30 kHz, respectively. Bottom panel: The reference input with the aforementioned frequencies. (c) Dynamical-decoupling-based lock-in detection at high frequencies. XY-4 sequence is shown as an example. The NV phase signal $\phi_{NV}$ is maximized when the detected field is in-phase with XY-4 (lower left plot), and minimized when out-of-phase by $\pi/2$ (lower right plot) (d) NV phase signals $\phi_{NV}$ at 200 kHz, 800 kHz and 1200 kHz. No discernible frequency dependence is observed. Pulse sequences see Supplementary Figure 6. (e) Frequency dependence of signal attenuation (top panel) and phase delay (bottom panel), extracted from the lock-in measurements in (b) and (d). Signal is normalized with respect to the expected phase induced by the applied electric field in the absence of screening. The red solid traces represent an ideal high-pass filter with a cut-off frequency of 35.4 kHz. PDD stands for periodic dynamical decoupling\cite{souza2012robust}, which refers to XY pulse sequences used at high frequencies. (f) A high-pass filter circuit model. NV is positioned within a metal gap. (g) A circuit model showing capacitive coupling between the diamond surface and a sample.} 
 \end{figure}

A strong electric field screening effect is observed at DC and low frequencies (see Supplementary Figure 10), likely caused by mobile charges on the diamond surface \cite{oberg2020solution}. Possible sources include adsorbed water layer under ambient conditions, electronic trapping states due to near-surface band-bending \cite{shirafuji1996electrical, stacey2019evidence}, and internal charge defects generated during diamond growth or NV creation \cite{collins2000spectroscopy}. We characterize this screening effect by measuring its frequency dependence. NV is positioned within a metal gap. An AC reference voltage is applied across the gap, generating an electric field at the NV. ``Lock-in" detection is employed to extract both the amplitude and phase of the NV signal relative to the reference signal. 

At low frequencies, we use a Ramsey-based lock-in sequence, where a train of equally spaced Ramsey measurements is synchronized with the reference signal, as shown in Fig.~\ref{fig:screening}a. In each Ramsey measurement, the NV spin is prepared in a superposition of $|0\rangle$ and $|+\rangle$, and then accumulates phase induced by the DC electric field within the free evolution time $\tau=800$ ns. The accumulated phase $\phi_{NV}$ can be extracted from the NV fluorescence signal measured after the final $\pi/2$ pulse (see Supplementary Note 2C). This phase is proportional to the local electric field, $\phi_{NV} = d_\perp E_{\perp} \cos{(2\phi_B+\phi_E)} \tau$, hence each Ramey measurement samples the electric field strength at the corresponding time step. The first Ramsey starts at 4 $\mu$s after the reference trigger, and the spacing between neighboring Ramsey measurements is also 4 $\mu$s. To have a sufficient number of sampling points, we performed measurements at reference signal frequencies below 50 kHz. As shown in Fig.~\ref{fig:screening}b, the NV Ramsey phase $\phi_{NV}$ oscillates in time, with the amplitude increasing with the reference frequency. Sinusoidal curve fitting extracts the amplitude and phase relative to the reference, which are represented by blue dots in Fig.~\ref{fig:screening}e.

At high frequencies, the lock-in detection is based on a dynamical-decoupling pulse sequence \cite{souza2012robust}, such as XY-4 shown in Fig.~\ref{fig:screening}c. The spacing between neighboring $\pi$ pulses is set to be half of the period of the reference signal. More $\pi$ pulses are inserted to match high frequencies. Due to the finite coherence time $T_2$, these measurements were performed above 200 kHz (see Supplementary Note 2A). The NV spin, prepared in a superposition of $|0\rangle$ and $|+\rangle$, accumulates a phase induced by the AC electric field within the evolution time $\tau$. The coherent phase signal $\phi_{NV}$ is maximized when the detected local field is `in-phase' $(\varphi=0)$ with the XY-4 sequence, and minimized when `90$\degree$ out-of-phase' $(\varphi=\pi/2)$. To measure the amplitude and phase of the detected signal relative to the reference, we vary the initial phase offset by sweeping the delay between the reference trigger and the first $\pi/2$ pulse. As shown in Fig.~\ref{fig:screening}d, $\phi_{NV}$ is maximized at zero initial phase offset, and minimized near $\pi/2$, which indicates very little relative phase between the detected and reference signal. In contrast to the low-frequency regime, here the oscillation amplitude shows no obvious dependence on the frequency. The extracted amplitude and phase by sinusoidal curve fitting are represented by orange dots in Fig.~\ref{fig:screening}e. 

Fig.~\ref{fig:screening}e summarizes the results in both low and high frequency regimes. The trend resembles the frequency response of a high-pass filter. The red solid curves represent an ideal high-pass filter with a cut-off frequency $f_c$ at 35.4 kHz, corresponding to an RC time constant $\sim 30$ $\mu s$. A simplified circuit model in Fig.~\ref{fig:screening}f illustrates a possible high-pass filter consisting of the resistive diamond surface and capacitance between the tip and electrodes. For a general sample, Fig.~\ref{fig:screening}g shows the capacitive coupling between the diamond surface and sample. Screening is significantly reduced at high frequencies due to the finite mobility of charge carriers. The specific frequency cut-off can vary between different probes, since the NV location, geometry of the tip surface and diamond purity all affect the screening effect. 

\subsection*{Spatial Mapping of AC and DC Electric Field Distribution} 
\begin{figure}[t]
\includegraphics[scale=0.55]{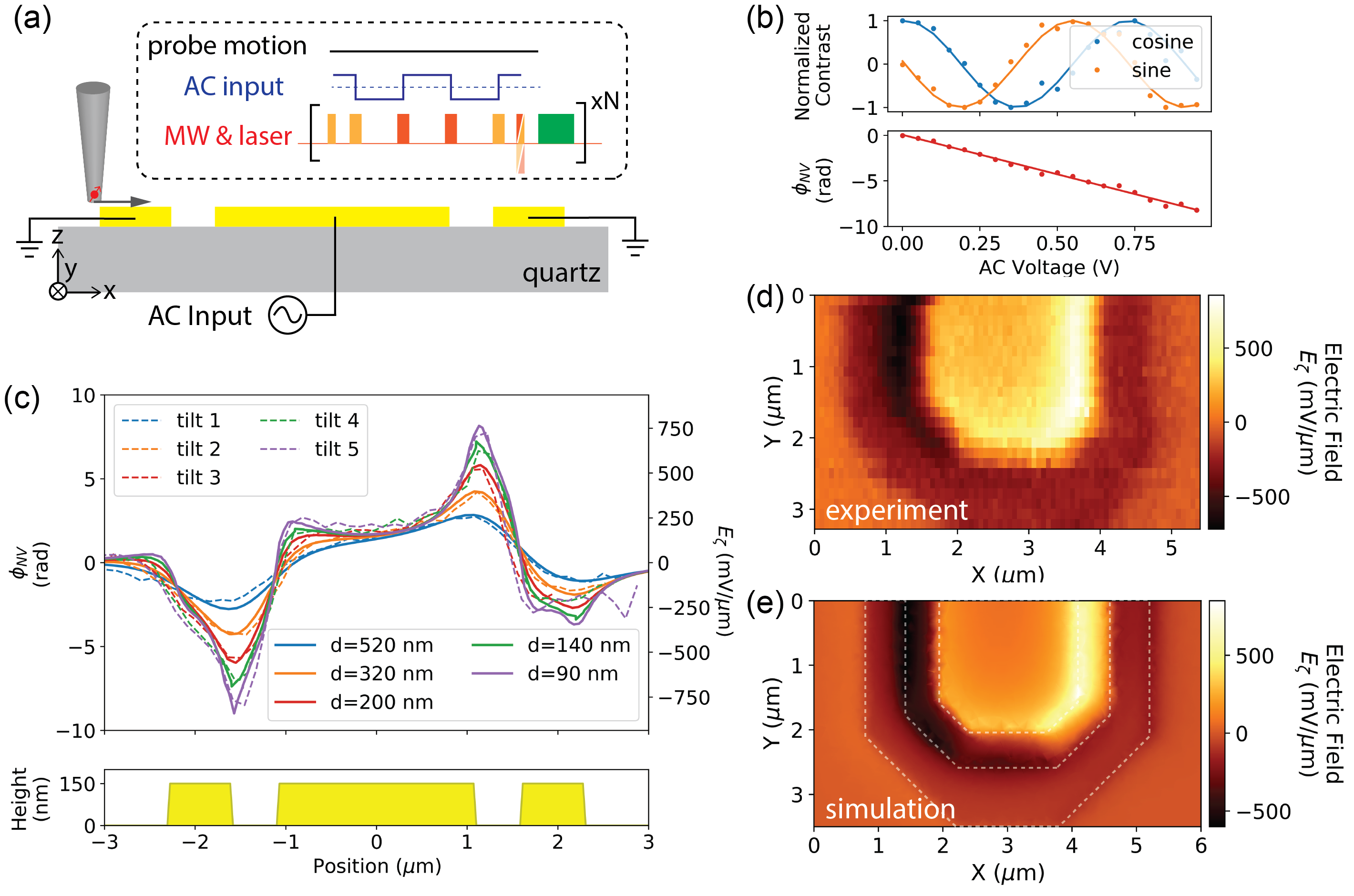} 
\caption{\label{fig:acimaging} AC Electric Field Imaging. (a) Schematic of scanning NV microscopy and the pulse sequence for AC electrometry. The probe oscillates with an amplitude $<$1 nm, hence represented by a black flat curve. The AC reference input is in-phase with the XY-4 detection sequence. (b) NV measurements performed at a fixed point within the gap. Top panel: the cosine and sine values measured by rotating the phase of the final $\pi/2$ pulse. Bottom panel: Extracted phase $\phi_{NV}$ proportional to the AC input amplitude. (c) Top panel: 1D line scans of the NV phase signal $\phi_{NV}$ at different NV-sample distances controlled by the tilting angle of the diamond probe. The AC input has a peak-to-peak amplitude of 0.96 V. Dashed traces are the experimental data, and solid traces show the COMSOL simulation. Bottom panel: the height profile of the metal. (d) 2D mapping of the AC electric field distribution in the U-shaped Au structure. (e) Simulated electric field distribution along the maximum coupling axis $\hat{\zeta}$ at a distance of 90 nm. The azimuth angle of $\hat{\zeta}$ with respect to the coordinate system shown in (a) is $\phi = 20 \degree$, and the zenith angle $\theta = 45 \degree$. The white dashed lines outline the edge of the Au structure used in the simulation.} 
\end{figure}

We now demonstrate imaging of the AC electric field distribution. A single NV at the apex of a diamond nanopillar (300 nm in diameter) scans over a U-shaped gold structure (Fig.~\ref{fig:acimaging}a). Our spatial resolution is limited by the NV-sample distance, which can be $\lesssim100$ nm. The diamond is attached to a piezoelectric tuning fork, which oscillates on resonance ($\sim$ 32 kHz) and provides frequency feedback to regulate the distance to the sample. In our AC electric field imaging, the oscillation amplitude is kept small (estimated to be $<$1 nm), and no reduction of the NV coherence time is observed. The probe motion is therefore represented by a flat line in Fig.~\ref{fig:acimaging}a. An AC signal $V_{pp}=0.96$ V at 250 kHz is applied to the middle electrode and synchronized in-phase with the XY-4 pulse sequence with a free evolution time $\tau=8$ $\mu s$. The accumulated phase is $\phi_{NV}=d_\perp \cdot E_\perp \cos{(2\phi_B+\phi_E)} \cdot  \tau = d_\perp E_\zeta \tau$, where $\hat{\zeta}$ denotes the maximum electric coupling direction, and $E_\zeta$ is the $\hat{\zeta}$ component of $E_\perp$. Fig.~\ref{fig:acimaging}b plots $\phi_{NV}$ as a function of the AC input amplitude, measured by the NV at a fixed point within the gap. $\phi_{NV}$ grows proportionally as expected. The cosine and sine values are measured by rotating the phase of the final $\pi/2$ pulse (see Supplementary Note 2C). Dashed traces in Fig.~\ref{fig:acimaging}c show 1D line scans of $\phi_{NV}$ and the corresponding e-field $E_\zeta$ at different NV-sample distances (controlled by the probe tilting angle). They are in good agreement with the simulated $E_\zeta$ distribution shown by the solid traces. The azimuth and zenith angles of $\hat{\zeta}$ in the simulation are at $\phi=20\degree$ and $\theta=45\degree$ respectively (Supplementary Figure 13a).  Based on a NV fluorescence rate $>$100 kcps, an optical contrast $>$20$\%$ and a phase accumulation time $\tau=8$ $\mu s$ in our experiment, we have achieved an AC electric field sensitivity of \SI{26}{\,mV/\mu m/\sqrt{Hz}} under ambient conditions (see Supplementary Note 2D). A 2D map of the AC electric field is shown in Fig.~\ref{fig:acimaging}d, and a simulated field distribution at a distance of d = \SI{90}{nm} is shown in Fig.~\ref{fig:acimaging}e. 

\begin{figure}[htbp!]
\includegraphics[scale=0.55]{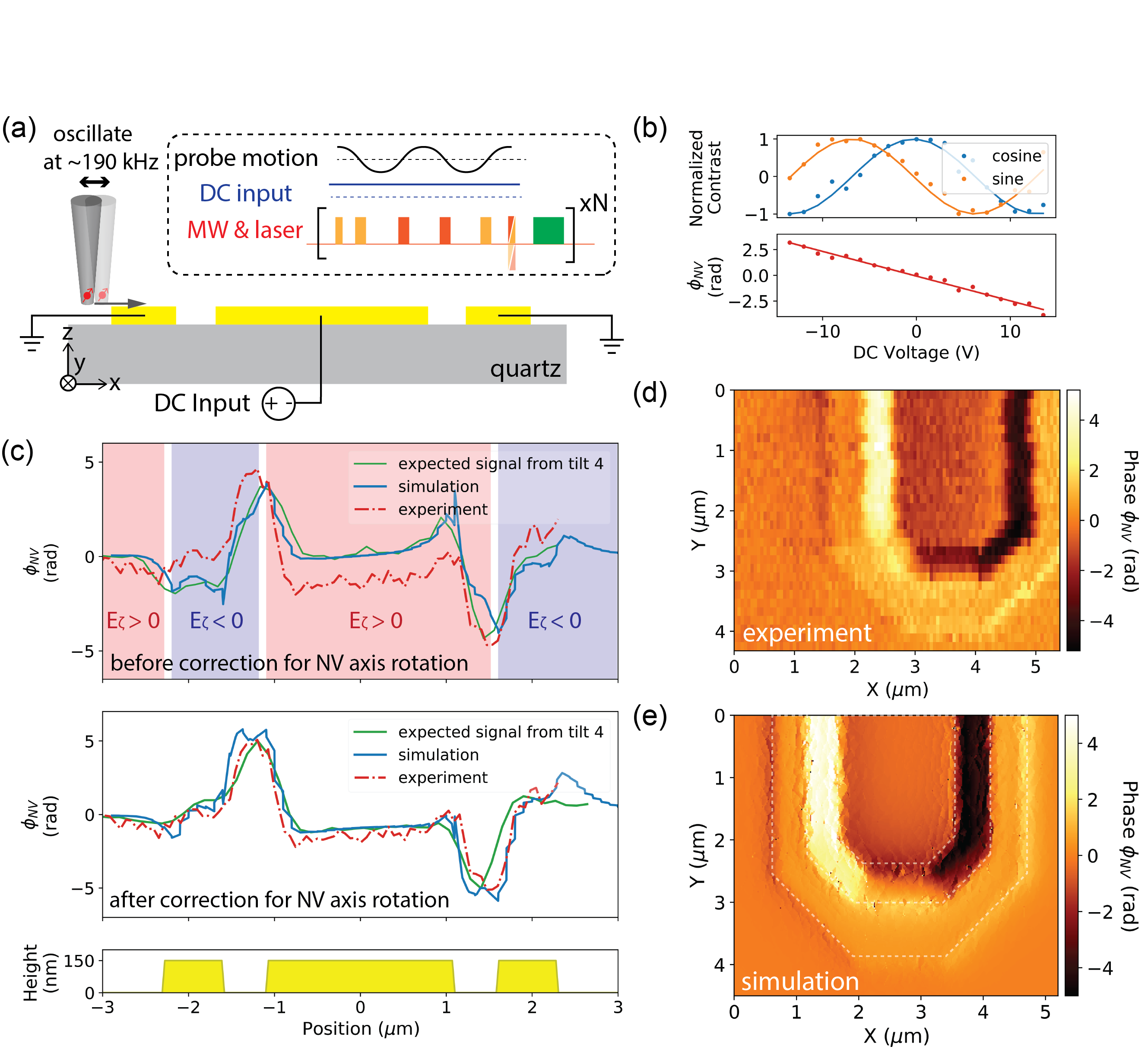} 
\caption{\label{fig:dcimaging} DC Electric Field Imaging. (a) Schematic of scanning NV microscopy and the pulse sequence for DC electrometry. The input signal is DC, represented by a blue flat line. The probe motion at $\sim$190 kHz is synchronized with the XY-4 sensing pulse sequences. (b) NV measurements performed at a fixed point within the gap. Top panel: the cosine and sine values measured by rotating the phase of the final $\pi/2$ pulse. Bottom panel: Extracted phase $\phi_{NV}$ proportional to the DC input amplitude. (c) 1D line scans of $\phi_{NV}$. The center electrode is at $V_{dc}=16$ V. Top panel: Discrepancy between the model $\phi_{NV} = d_\perp E_{\zeta}'(x) A \tau$ and data. The green trace shows the expected gradient field signal calculated from tilt 4 in Fig.~\ref{fig:acimaging}c. The blue trace shows the COMSOL simulation. The red (blue) background corresponds to the region of $E_{\zeta}>0$ ($E_{\zeta}<0$) as shown in Fig.~\ref{fig:acimaging}c. Middle panel: The model $\phi_{NV} = d_\perp (E_{\zeta}'(x) A + \beta E_{\zeta}) \tau$ showing an improved agreement with data, where $A$ = 13 nm is the oscillation amplitude and $\beta$ = -0.03 is the proportionality constant that captures the NV axis rotation. Bottom panel: the height profile of the metal. (d) 2D mapping of $\phi_{NV}$ measured at $V_{dc}$ = 16 V. (e) Simulated signal $\phi_{NV}$ at a distance of 140 nm from the sample. The maximum coupling axis $\hat{\zeta}$ is set to be same as in Fig.~\ref{fig:acimaging}e, and the oscillation amplitude is \SI{\,13}{\,nm}. The white dashed lines outline the edge of the Au structure used in the simulation.} 
 \end{figure}

To map DC electric fields, we employ a motion-enabled imaging technique \cite{hong2013nanoscale} that converts the DC signal to AC in order to overcome the screening effect. More concretely, the probe oscillates with a relatively large amplitude ($>10$ nm) such that the NV experiences an AC local field proportional to the local spatial gradient $E_{\zeta}'(x)$ (Fig.~\ref{fig:dcimaging}a). In addition, the bending of the tuning fork induces a rotational motion of the NV axis, giving rise to an AC modulation proportional to the local static field $E_{\zeta}$. This latter effect cannot be ignored as shown by the data below. The amplitude of the total motion-enabled AC signal has a form of $E_{amp}=E_{\zeta}'(x) A + \beta E_{\zeta}$, where $A$ is the probe oscillation amplitude and $\beta$ is an empirical constant capturing the NV axis rotation. To operate at a sufficiently high frequency at which the screening diminishes, we drive the `clang' mode of the tuning fork at $\sim190$ kHz \cite{rossing1992acoustics} (Supplementary Figure 11). The sensing pulse sequence is synchronized with the mechanical motion, so the NV accumulates a coherent phase $\phi_{NV} = d_\perp E_{amp} \tau$. Fig.~\ref{fig:dcimaging}b shows the NV measurement at a fixed point within the gap, where $\phi_{NV}$ is proportional to the applied DC voltage $V_{dc}$. Based on our experimental parameters, we achieved a DC field gradient sensitivity of \SI{2}{\,V/\mu m^2/\sqrt{Hz}} (Supplementary Note 2D). Fig.~\ref{fig:dcimaging}c compares a 1D line scan of $\phi_{NV}$ along the $\hat{x}$-axis at $V_{dc}=16$ V with the expected signal deduced from the measurements in Fig.~\ref{fig:acimaging}c and simulated signal by COMSOL. The top panel shows a clear discrepancy between data and expected signal or simulation, when only the gradient term $E_{\zeta}'(x)A$ in $E_{amp}$ is considered. The sign of the discrepancy coincides with the sign of $E_\zeta$ shown in Fig.~\ref{fig:acimaging}c. Including the term $\beta E_{\zeta}$ in $E_{amp}$ leads to an improved agreement between data and model (with $A$ = 13 nm and $\beta$ = -0.03), as shown in the middle panel. Since the probe oscillates at a large amplitude while performing contact-mode AFM scanning, the actual motion highly depends on the details of probe-sample engagement. This is challenging to simulate accurately and could account for the remaining discrepancy. A 2D map of $\phi_{NV}$ is shown in Fig.~\ref{fig:dcimaging}d, showing a reasonable agreement with the simulation result in Fig.~\ref{fig:dcimaging}e. 

\section*{Discussion}
In conclusion, we demonstrated electric field imaging with a single NV at the apex of a diamond scanning tip under ambient conditions. The diamond surface significantly screens external electric fields at low frequencies. We quantitatively measured its frequency dependence using `lock-in' detection sequences, and therefore revealed the RC time constant of the tip surface ($\sim$ \SI{\,30}{\,\mu s}). To overcome screening, we introduced a motion-enabled technique and demonstrated spatial mapping of the DC electric field gradients. 

Potential improvements to these demonstrations can be achieved as follows. First, AFM operating in tapping mode would avoid direct contact between the probe and sample and allow the probe to oscillate with a larger amplitude more stably. This will give a higher field gradient sensitivity in motion-enabled imaging. Second, given an unknown sample, the DC field distribution along the maximum coupling axis $\zeta$ can be reconstructed by carefully pre-calibrating the probe oscillation direction and amplitude using a well-defined sample as described in this paper. Third, using multiple NVs of different orientations or multiple pillars with NVs of different orientations, one can extract both the magnitude and the direction of external electric fields, i.e. vector electrometry \cite{yang2020vector}. Finally, a higher spatial resolution is achievable by using even shallower NVs and motorized goniometers for a better control of the probe tilting angle. 

NV electrometry possesses a unique combination of properties as compared to other existing techniques. In this work, we achieved an AC electric field sensitivity of \SI{26}{\,mV/\mu m/\sqrt{Hz}}, DC electric field gradient sensitivity of \SI{2}{\,mV/\mu m^2/\sqrt{Hz}}, and sub-100 nm spatial resolution limited by the NV-sample distance, all under ambient conditions. Most of other electrometry techniques are based on measuring potentials and many require cryogenic conditions. For example, scanning single-electron-transistor (SET) \cite{yoo1997scanning, yacoby1999electrical, martin2008observation, sulpizio2019visualizing} is capable of measuring microvolt local potential, however this remarkable sensitivity requires low-temperature operation and its spatial resolution is limited by the tip size ($>$100 nm). Kelvin probe force microscopy (KPFM) \cite{nonnenmacher1991kelvin, melitz2011kelvin} and electrostatic force microscopy (EFM) \cite{girard2001electrostatic} can achieve sub-10nm spatial resolution and operate under ambient conditions, however they are not optimal for quantitative electric field measurements. NV center is therefore a valuable addition as a unique electric field sensor with complementary advantages. We also highlight that NVs have unparalleled sensing versatility with a broad operating frequency range. Integrating electrometry and magnetometry into a single scanning probe will open up exciting opportunities in imaging nanoscale phenomena. 

\section*{Methods}
\subsection*{Diamond scanning probe and AFM control} 
The diamond probe was made from an electronic-grade CVD diamond purchased from Element Six, with a natural abundance (1.1$\%$) of $^{13}$C impurity spins. 
The probe is of $\sim$ \SI{50}{\,\mu m} $\times$ \SI{55}{\,\mu m} $\times$ \SI{125}{\,\mu m} in dimension. Each probe has seven tips in a row with a spacing of \SI{7}{\,\mu m}. Details of the multiple-pillar probe design and fabrication process are described in \cite{zhou2017scanning, xie2018crystallographic, vool2021imaging}. 

The probe is attached to one prong of a quartz tuning fork using optical adhesive (Thorlabs NOA63). Two manual goniometers (Edmund) control its tilting angle and the NV-sample distance. The AFM contact between probe and sample is controlled by an attocube SPM controller (ASC500). Piezoelectric nanopositioners (attocube ANPxyz101 and ANPxyz100) were used for sample scanning. More details can be found in Supplementary Note 2E.

\subsection*{Optical setup} 
NV experiments were performed on a home-built confocal laser scanning microscope. A 532 nm green laser (Cobolt Samba 100), focused by a 100$\times$, NA=0.7 objective (Mitutoyo M Plan NIR HR), was used to initialize the NV spin to the $|m_S=0\rangle$ state and generate spin-dependent photoluminescence for optical readout.  An avalanche photodiode (APD, Excelitas Technologies Photon Counting Module SPCM-AQRH-13) was used to measure the NV fluorescence rate. 

\subsection*{Quantum control setup} 
The microwaves were generated from a signal generator (Rohde $\&$ Schwarz SGS100A 6GHz SGMA RF Source) and amplified by a MW amplifier (Amplifier Research 30S1G6). All quantum measurements were performed on the Quantum Orchestration Platform (Quantum Machines). An Operator-X (OPX) generated control pulse sequences, output AC input voltages ($V_{pp}$ = 0.96 V), and measured the photon counts. DC input ($V_{dc}$ = 16V) is provided by a voltage source (Yokogawa GS200). 

\subsection*{Electrostatics simulation} 
The finite-element calculation package COMSOL performs the electrostatic simulations. To model the geometry, we use real device dimensions by importing the lithography design into the software. 2D simulation produces plots in Fig.~\ref{fig:acimaging}c and Fig.~\ref{fig:dcimaging}c. 3D simulation produces the plots in Fig.~\ref{fig:acimaging}e and Fig.~\ref{fig:dcimaging}e. 

\def\bibsection{\section*{References}} 
\bibliographystyle{naturemag}
\bibliography{esensing_reference}

\section*{Data Availability}
All data generated and analyzed during the study are included in this article and supplementary information.

\section*{Acknowledgments}
We thank Yonatan Cohen and Niv Drucker for their support on the Quantum Orchestration Platform, Shaowen Chen and Elizabeth Park for their assistance in the sample fabrication and diamond probe gluing processes. This work was primarily supported by the Quantum Science Center (QSC), a National Quantum Information Science Research Center of the U.S. Department of Energy (DOE). 

\section*{Author Contributions}
Z.Q. and A.Y. conceived the project. Z.Q. fabricated the sample, performed the measurements, simulation, and analyzed the data. Z.Q., A.H., U.V., and A.Y. discussed the results. T.X.Z. fabricated the diamond scanning probes. Z.Q. and T.X.Z. built the scanning probe setup. A.Y. supervised the project. All the authors contributed to the writing of the manuscript.

\section*{Competing Interests}
The authors declare no competing interests.

\end{document}

\title{Supplementary Information for \\ Nanoscale Electric Field Imaging with an Ambient Scanning Quantum Sensor Microscope}

\author{Ziwei Qiu}
\email[ziweiqiu29@gmail.com]{}
\affiliation{Department of Physics, Harvard University, Cambridge, MA 02138, USA}
\affiliation{John A. Paulson School of Engineering and Applied Sciences, Harvard University, Cambridge, MA 02138, USA}
\author{Assaf Hamo}
\email[Present address: Department of Physics, Bar-Ilan University, Ramat Gan, Israel]{}
\affiliation{Department of Physics, Harvard University, Cambridge, MA 02138, USA}
\author{Uri Vool}
\email[Present address: Max Planck Institute for Chemical Physics of Solids, Dresden, Germany]{}
\affiliation{Department of Physics, Harvard University, Cambridge, MA 02138, USA}
\affiliation{John Harvard Distinguished Science Fellows Program, Harvard University, Cambridge MA 02138, USA}
\author{Tony X. Zhou}
\email[Present address: Research Laboratory of Electronics, Massachusetts Institute of Technology, Cambridge, MA, USA]{}
\affiliation{Department of Physics, Harvard University, Cambridge, MA 02138, USA}
\affiliation{John A. Paulson School of Engineering and Applied Sciences, Harvard University, Cambridge, MA 02138, USA}
\author{Amir Yacoby}
\email[Correspondence to: yacoby@physics.harvard.edu]{}
\affiliation{Department of Physics, Harvard University, Cambridge, MA 02138, USA}
\affiliation{John A. Paulson School of Engineering and Applied Sciences, Harvard University, Cambridge, MA 02138, USA}

\date{\today}
\maketitle
\tableofcontents 

\section{I. NV electrometry under a perpendicular magnetic field}
\subsection{A. Electron spin eigenstates}
\begin{figure}[htbp!]
\centering
\includegraphics[scale=0.55]{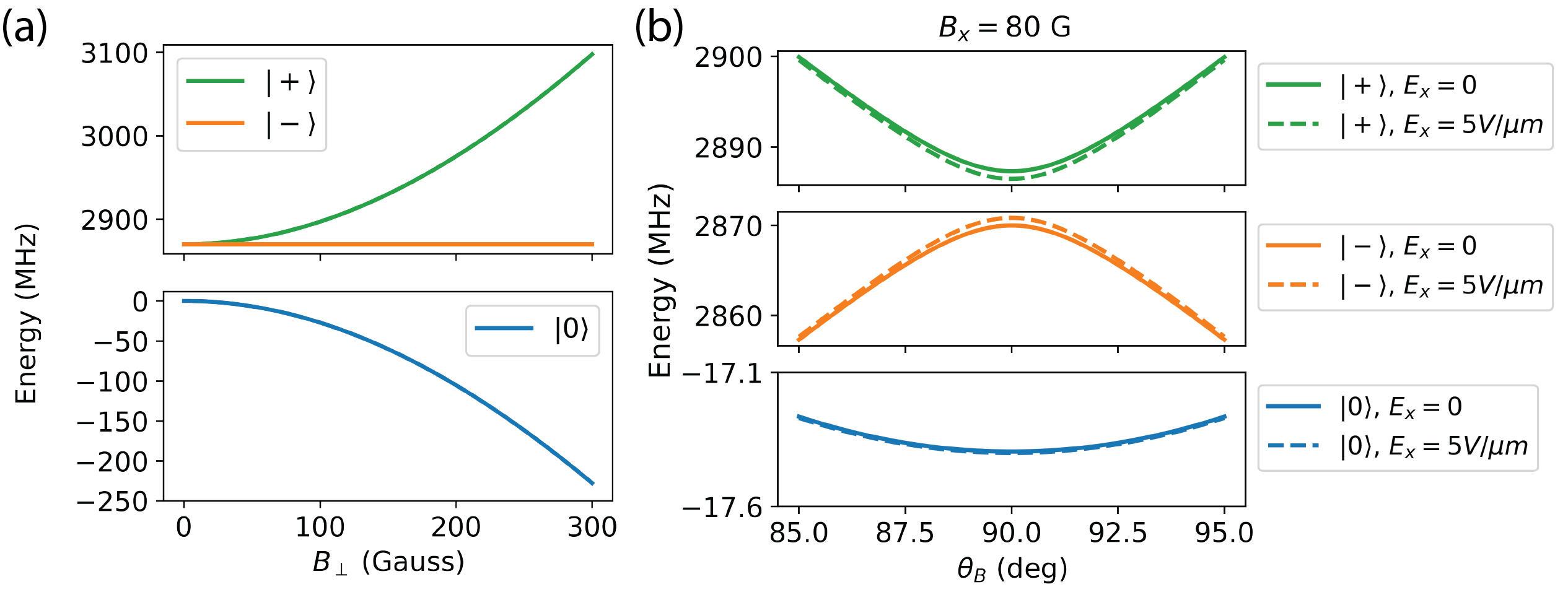}
 \caption{\label{fig:energy_level_no_nuclear} The electron eigenstates $|0, \pm \rangle$ under a perpendicular magnetic field. (a) Eigenstate energy levels as a function of the perpendicular field magnitude. (b) Stark shifts induced by electric fields. Both the magnetic and electric fields are assumed to be along $\hat{x}$ in producing the plot. The coordinate system is the same as defined in Figure 1b of the main text.  }
 \end{figure}

Under a bias magnetic field that is perpendicular to the NV axis, the NV electron spin ground state Hamiltonian is $H_e=D_{gs}S_z^2+\gamma_B (B_xS_x+B_yS_y)$, where $D_{gs}\approx$ 2.87 GHz is the zero-field splitting (ZFS) and $\gamma_B$ = 2.8 MHz/G is the electron spin gyromagnetic ratio. The eigenstates, represented by $|0, \pm\rangle$, are superpositions of $|m_S=0,\pm1\rangle$: $|0\rangle \approx |m_S = 0\rangle$, $|\pm \rangle \approx \frac{1}{\sqrt{2}}(|m_S=1\rangle + e^{2i\phi_B}|m_S=-1\rangle)$\cite{dolde2011electric}. As depicted in Figure 1d of the main text, the energy levels of $|\pm\rangle$ are sensitive to electric field perturbations in the XY plane $E_\perp$, with a coupling strength $d_\perp =$ \SI[separate-uncertainty=true]{0.17\pm 0.03}{\,MHz/V/\mu m} (Figure~\ref{fig:energy_level_no_nuclear}). Coupling to the electric field along the NV axis $E_z$ is negligible, where the coupling strength is $d_\parallel=$ \SI[separate-uncertainty=true]{3.5\pm0.2}{\times10^{-3}\,MHz/V/\mu m}, so it is ignored in our analysis. We also ignore the strain in the diamond. 

\subsection{B. $^{15}$NV vs. $^{14}$NV}
We choose to work with $^{15}$NV in our experiments. $^{15}$N has nuclear spin $I=1/2$. The hyperfine interaction between the nuclear spin and electron spin is highly sensitive to the magnetic field angle, giving rise to electron spin echo envelope modulation (ESEEM). The magnetic field angle is carefully calibrated such that the electron spin coherence is maximized \cite{qiu2021nuclear}. 

The different nuclear spins carried by $^{15}$N and $^{14}$N cause subtle and important differences in implementing electric field sensing, as summarized in Table~\ref{table:comparison}. For $^{15}$N, nuclear spin is $I=1/2$, and the hyperfine coupling constant is $A_{||}$=3.65 MHz, $A_{\perp}$=3.03 MHz. For $^{14}$N, nuclear spin is $I=1$, and the hyperfine coupling constant is $A_{||}=A_{\perp}$=2.2 MHz \cite{gali2008ab, felton2009hyperfine}. Both require a (nearly) perpendicular bias magnetic field for electric field sensing. The full NV Hamiltonian is $H = D_{gs}S_z^2+\gamma_B\vec{B}\cdot\vec{S}+\vec{I}\cdot\vec{A}\cdot\vec{S}+\gamma_N\vec{B}\cdot\vec{I}+d_{||}E_zS_z^2+d_{\perp}E_x(S_y^2-S_x^2)+d_{\perp}E_y(S_xS_y+S_yS_x)$. $^{14}$NV has an additional term due to the quadrupole interaction, $QI_z^2$, where $Q\approx$-5.01 MHz. Under a perpendicular magnetic field, the energy splitting between $|+\rangle$ and  $|-\rangle$ is roughly $\gamma_B^2B_{\perp}^2/D_{gs}$. We discuss two magnetic field regimes.

\begin{table}[htbp!]
\caption{$^{14}$NV electrometry vs. $^{15}$NV electrometry} 

\centering
\begin{tabular}{l*{6}{c}r}
B Field              & $^{14}$NV (I=1) & $^{15}$NV  (I=1/2) \\
\hline
Weak B & \makecell{only one out of three $m_I$ states is \\sensitive to electric field}  &  \makecell{only one out of two $m_I$ states is \\sensitive to electric field} \\
Strong B & \makecell{driving efficiency is \\nuclear spin dependent} & \makecell{both nuclear spin states contribute to signal\\ and driving efficiency is nuclear spin independent}    \\
\hline
\end{tabular}
\label{table:comparison}
\end{table}

\begin{figure}[b]
\centering
\includegraphics[scale=0.5]{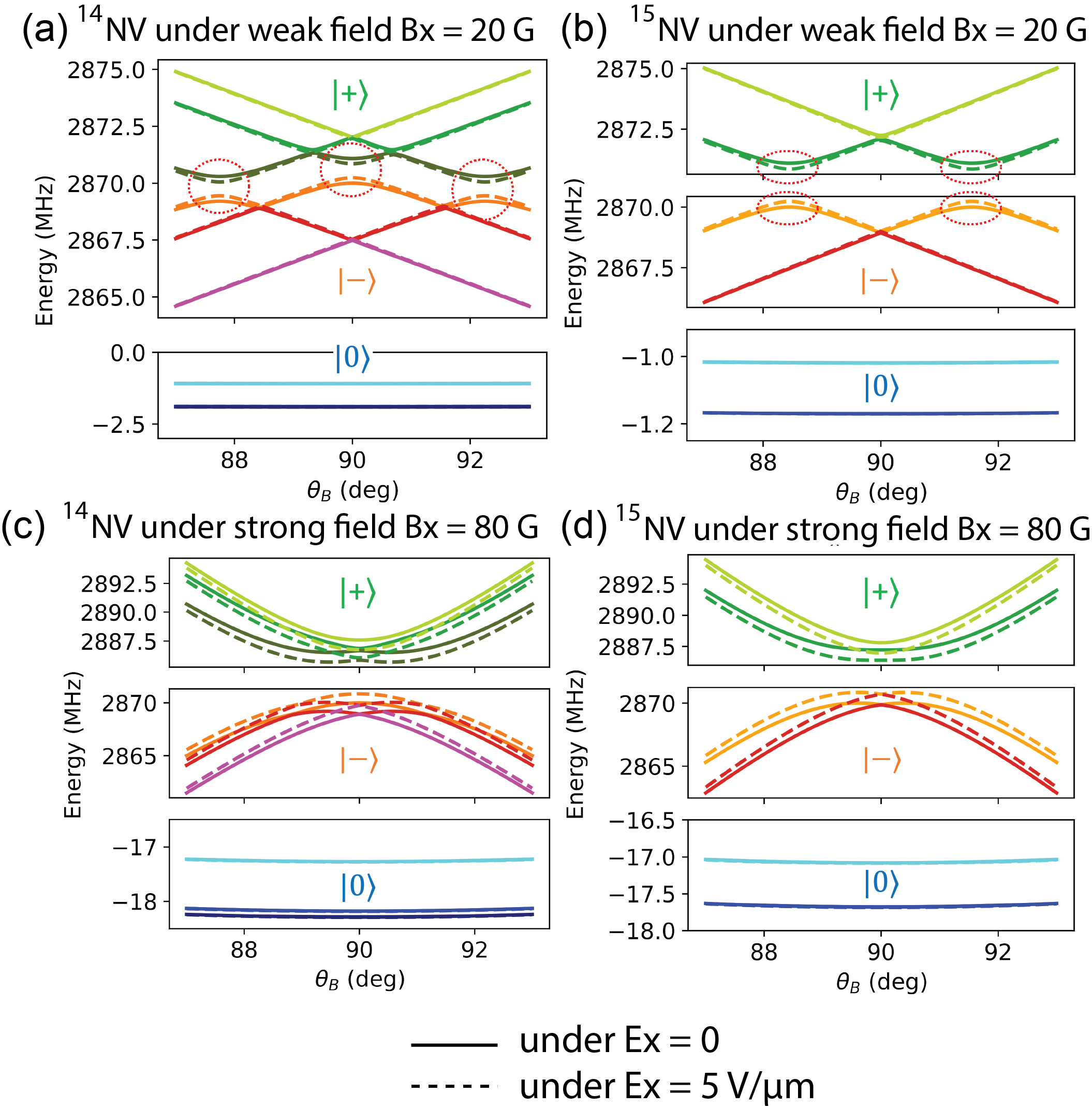}
 \caption{\label{fig:nuclear_spin_effect} Electric field sensing with $^{14}$NV and $^{15}$NV under different magnetic fields. (a)(b) Under a weak field, only one nuclear spin state is sensitive to electric fields. (c)(d) Under a strong field, all nuclear spin states are sensitive to electric fields.}
 \end{figure}
 
When $\gamma_B^2B_{\perp}^2/D_{gs} < A_{\perp, ||}$, i.e. $B_{\perp} <$ 33 G, the hyperfine splitting is greater than the $|\pm\rangle$ splitting. At $\theta_B=90\degree$, where $\theta_B$ is the angle between $\vec{B}$ and NV axis, only the $m_I=0$ state is sensitive to magnetic field, as shown in Figure~\ref{fig:nuclear_spin_effect}a. Intuitively, for $m_I\neq0$ states, the nuclear spin exerts a hyperfine field at the electron spin, so the total magnetic field seen by the electron spin is no longer perpendicular to the NV axis. The external bias magnetic field needs to be slightly off from $\theta_B=90\degree$ in order for the $m_I\neq0$ states to be electric field sensitive, as shown in Figure~\ref{fig:nuclear_spin_effect}a for $^{14}$NV and Figure~\ref{fig:nuclear_spin_effect}(b) for $^{15}$NV. For electric field sensing to work in this weak magnetic field regime, the NV dephasing time $T_2^*$ needs to be sufficiently long to resolve the nuclear spin sublevels. In addition, the NV fluoresence contrast, hence the signal-to-noise ratio, is reduced by $\frac{2}{3}$ for $^{14}$NV and $\frac{1}{2}$ for $^{15}$NV.

When $\gamma_B^2B_{\perp}^2/D_{gs} > A_{\perp, ||}$, i.e. $B_{\perp} >$ 33 G,  the $|\pm\rangle$ splitting is greater than the hyperfine splitting, and all the $m_I$ states are sensitive electric field. This is because the hyperfine field that the nuclear spin exerts on the electron spin is much smaller than the external bias field, $\theta_B\approx90\degree$ holds for all nuclear spin states. As shown in Figure~\ref{fig:nuclear_spin_effect}c and d, a 5 $V/\mu m$ electric field causes energy shifts on all the nuclear spin states. 

\begin{figure}[htbp!]
 \centering
\includegraphics[scale=0.45]{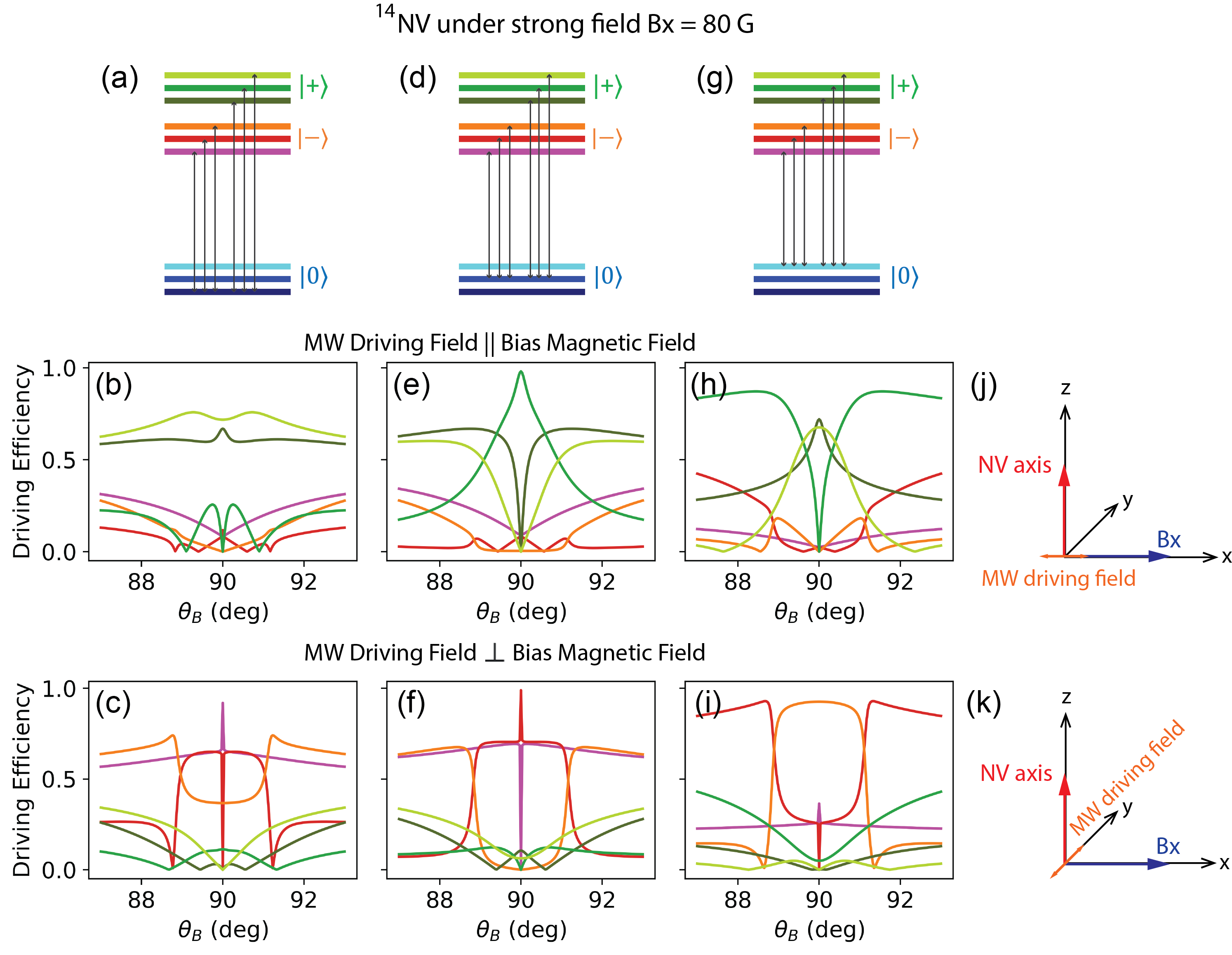}
 \caption{\label{fig:driving_efficiency_14NV} Transition efficiencies of $^{14}$NV under $B_x$=80 G. (a)(b)(c) Transition efficiencies between the lowest $|0\rangle$ nuclear sublevel and all upper levels.  (b) plots the MW driving efficiency when the MW field is along the magnetic field direction, as shown in (j). (c) plots the MW driving efficiency when the MW field is perpendicular to the magnetic field direction, as shown in (k). Similarly, (d)(e)(f) and (g)(h)(i) plot the transition efficiencies between the middle and highest $|0\rangle$ nuclear sublevel and upper levels. }
 \end{figure}

\begin{figure}[htbp!]
\centering
\includegraphics[scale=0.45]{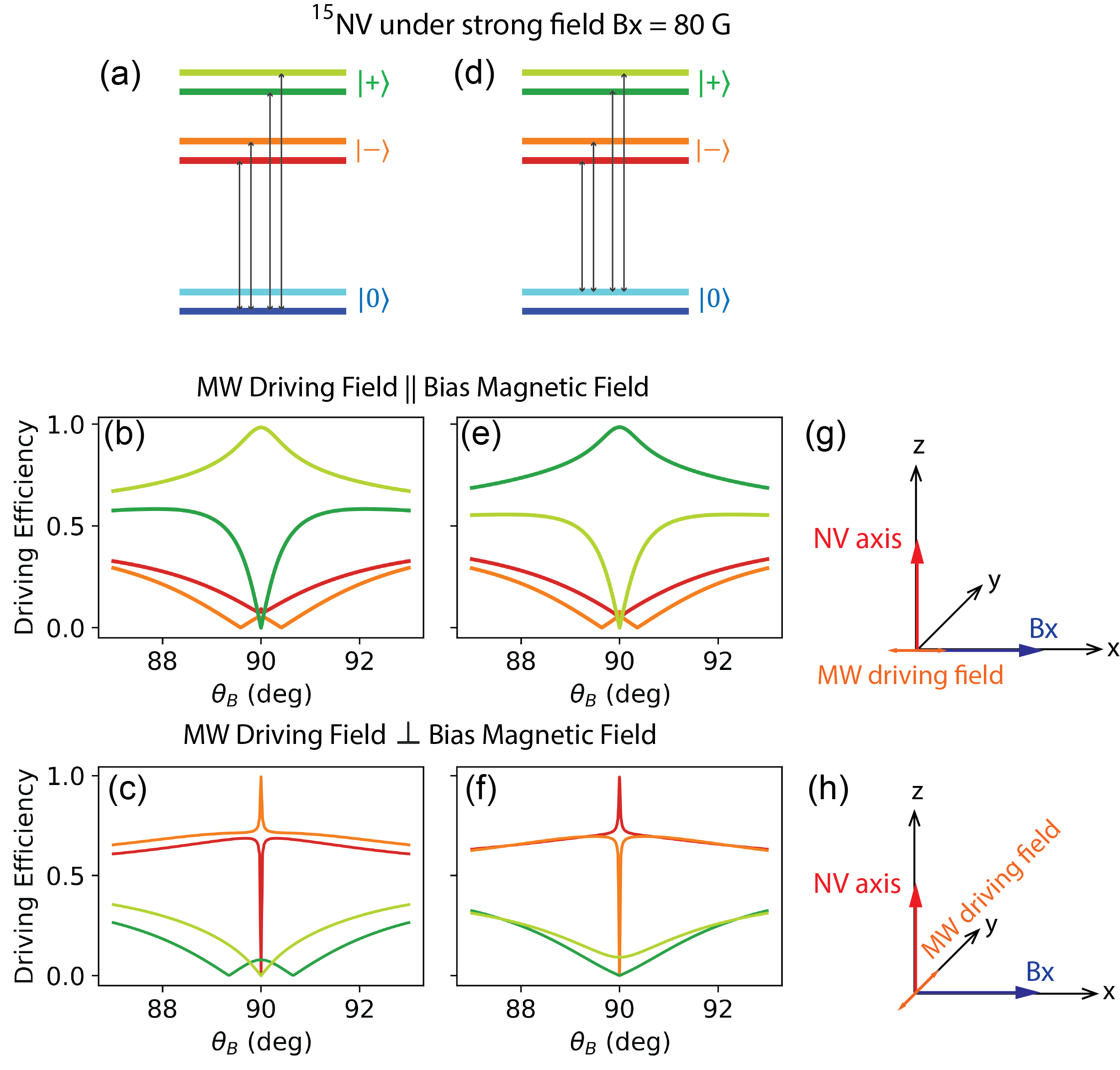}
\caption{\label{fig:driving_efficiency_15NV} Transition efficiencies of $^{15}$NV under $B_x$=80 G. (a)(b)(c) Transition efficiencies between the lower $|0\rangle$ nuclear sublevel and all upper levels.  (b) plots the MW driving efficiency when the MW field is along the magnetic field direction, as shown in (g). (c) plots the MW driving efficiency when the MW field is perpendicular to the magnetic field direction, as shown in (h). Similarly, (d)(e)(f) plots the transition efficiencies between the second $|0\rangle$ nuclear sublevel and upper levels. }
 \end{figure}

However, in the strong B regime, there is another subtle difference between $^{15}$NV and $^{14}$NV due to microwave (MW) driving. The MW field component parallel (perpendicular) to the bias magnetic field efficiently drives the transition between $|0\rangle$ and $|+\rangle$ ($|-\rangle$). Normally, MW field drives the transition between electron states while the nuclear spin state remains unchanged. However under a nearly perpendicular magnetic field, the nuclear spin state becomes electron spin dependent \cite{qiu2021nuclear}. Transitions between all pairs of states can become allowed to some degree, and the transition amplitude is highly angle dependent and nuclear spin dependent. For example, between $|0\rangle$ and $|+\rangle$ in $^{14}$NV, there are in total $3\times3=9$ possible transitions. Consequently, there is no well-defined $\pi$ pulse duration. After applying a MW pulse, the electron spin can end up in different transversal plans in the Bloch Sphere, which results in a short $T_2^*$ in measuring Rabi oscillation. Comparing Figure~\ref{fig:driving_efficiency_14NV} and Figure~\ref{fig:driving_efficiency_15NV}, this problem is more severe in $^{14}$NV than in $^{15}$NV. As shown in Figure~\ref{fig:driving_efficiency_15NV}c and d, when driving transitions between $|0\rangle$ and $|+\rangle$ and $\theta_B\approx90\degree$, only one transition is allowed for each nuclear spin sublevel. 

\section{II. Experimental details}
\subsection{A. Sample and NV Characteristics}
The U-shaped gold structure and the RF stripline were made of Au/Ti (140nm / 12nm in thickness) by thermal evaporation on a quartz substrate. The RF stripline is about \SI{20}{\,\mu m} in width.

The diamond probe was made from an electronic-grade CVD diamond, with a natural abundance (1.1$\%$) of $^{13}$C impurity spins. NV centers are created by $^{15}$N ion implantation followed by vacuum annealing. They are estimated to be $<$40 nm deep from the surface. Details of probe fabrication can be found in \cite{zhou2017scanning}. 

\begin{figure}[htbp!]
\centering
\includegraphics[scale=0.65]{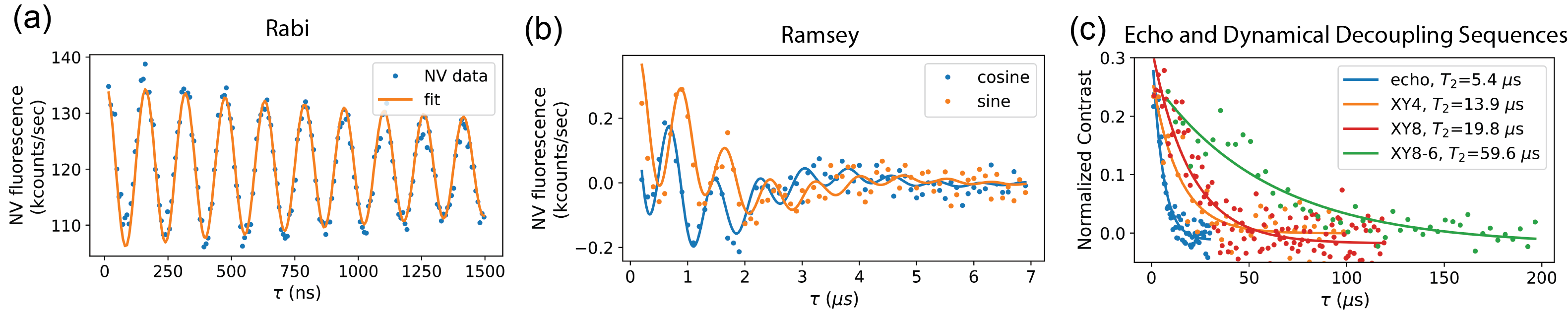}
\caption{\label{fig:NV_characterization} Characterizations of NV coherent properties. (a) Rabi oscillation between $|0\rangle$ and $|+\rangle$. (b) Ramsey fringes measured with a detuning of $\sim$1 MHz. (c) Coherence decay measured by spin-echo and dynamical decoupling sequences.}
\end{figure}

\begin{figure}[htbp!]
\centering
\includegraphics[scale=0.45]{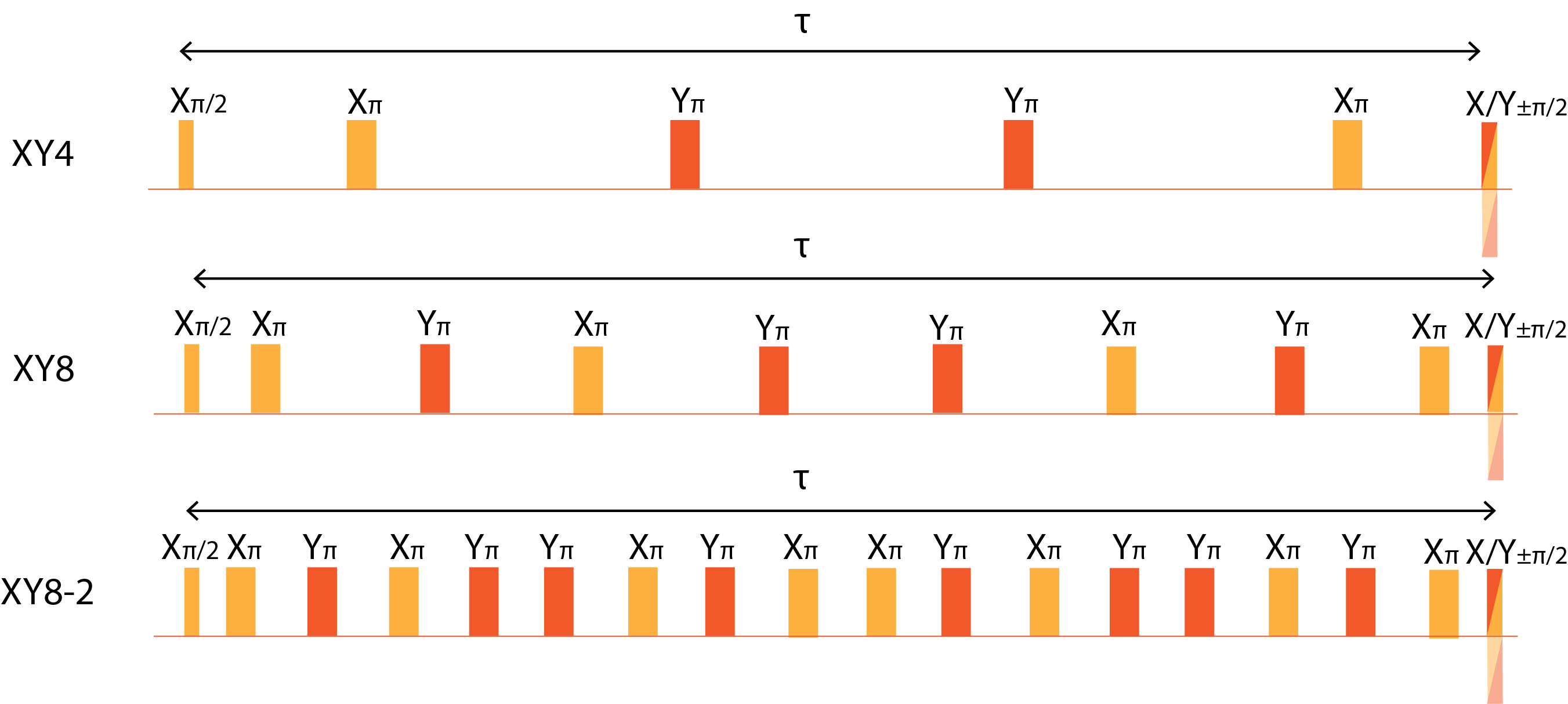}
\caption{\label{fig:DD_seq} Dynamical decoupling pulse sequences \cite{souza2012robust}.}
\end{figure}

We characterized the NV under a bias magnetic field $\sim$100 G oriented perpendicular to the NV axis. As shown in Figure 1e of the main text, due to the microwave field polarization, the $|0\rangle$-to-$|+\rangle$ transition is more efficiently driven than the $|0\rangle$-to-$|-\rangle$ transition. Figure~\ref{fig:NV_characterization}a shows the Rabi oscillation between $|0\rangle$ and $|+\rangle$. Figure~\ref{fig:NV_characterization}b shows the Ramsey fringes measured with a detuning of $\sim$1 MHz relative to the $|0\rangle \leftrightarrow |+\rangle$ transition frequency. The extracted dephasing time $T_2^*$ is about \SI{1.5}{\,\mu s}. We also measured the coherence time of the $|0\rangle$-$|+\rangle$ superposition state using spin-echo and dynamical decoupling sequences  with multiple equally spaced $\pi$ pulses (Figure~\ref{fig:DD_seq})\cite{souza2012robust}. As shown in Figure~\ref{fig:NV_characterization}c, $T_2$ increases as the number of $\pi$ pulses. 

Lock-in detection based on dynamical decoupling sequences was used to characterize the screening effect at high frequencies (see Figure 2c-d of the main text). Due to the finite coherence time shown in Figure~\ref{fig:NV_characterization}c, measurements were performed above 200 kHz. A XY-4 sequence with $\tau$ = \SI{8}{\mu s} was used in producing the data shown in Figure 3 and Figure 4 of the main text.

\subsection{B. Experimental setup and control} 

Experiments were performed on a home-built confocal laser scanning microscope, with an AFM scanning capability \cite{maletinsky2012robust}. The AFM contact between probe and sample is controlled by an attocube SPM controller (ASC500). Piezoelectric nanopositioners (attocube ANPxyz101 and ANPxyz100) were used for sample scanning. The quantum measurements were performed on the Quantum Orchestration Platform, a universal quantum control platform developed by Quantum Machines. An Operator-X (OPX) generated control pulse sequences and measured the photon counts. Microwaves are generated by RhodeSchwartz SGS100A and amplified (Amplifier Research Model 30S1G6). NVs are optically addressed with an NA = 0.7 air objective from above, as depicted in Figure 1a of the main text. 532 nm laser of power $\sim$1 mW, pulsed by an acoustic-optical modulator (AOM), was used to initialize and readout the NV spin states. An avalanche photodiode (APD) collects the red photons from the NV fluorescence and sends the signal to the OPX.

\subsection{C. Electric field sensing sequence and phase extraction}

\begin{figure}[htbp!]
\centering
\includegraphics[scale=0.5]{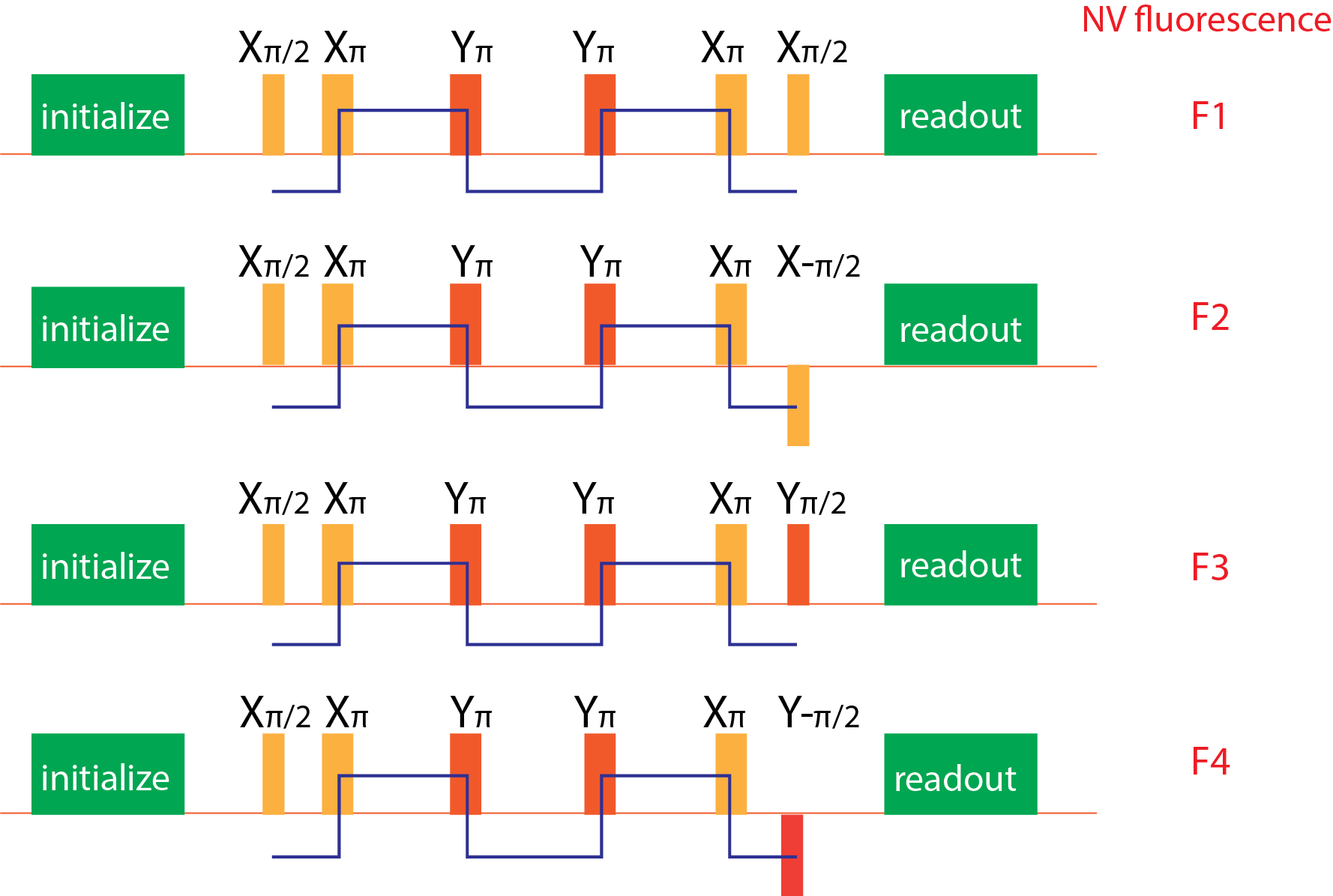}
\caption{\label{fig:sequence} AC electric field sensing sequence. The AC electric field signals, shown by the blue lines, are synchronized with the microwave pulses.}
\end{figure}

Figure~\ref{fig:sequence} illustrates the XY-based AC sensing sequence used to extract the phase signal $\phi_{NV}$. The AC electric field is synchronized with the MW pulse, as shown by the blue curves. The cosine and sine values of $\phi_{NV}$ are measured by rotating the phase of the final $\frac{\pi}{2}$ pulse:
\begin{eqnarray}
\cos{\phi_{NV}} = 2 \cdot \frac{F_2-F_1}{F_2+F_1}, \\
\sin{\phi_{NV}} = 2 \cdot \frac{F_4-F_3}{F_4+F_3},
\end{eqnarray}
where $F_1$, $F_2$, $F_3$ and $F_4$ are the NV fluorescence rates measured by the green laser pulse at the end the four sequences in Figure~\ref{fig:sequence}, respectively. 

\subsection{D. Electric field sensitivity}
We first analyze the AC electric field sensitivity $\eta_{E}$ detected by the sequence as shown in Figure 3a of the main text. The sensitivity is determined by the NV fluorescence count rate $F$ (unit: counts per second), the optical contrast $C$ between the states $|0\rangle$ and $|+\rangle$, the total phase accumulation time $\tau$, the APD readout window $T_r$, initialization time $t_{ini}$ as well as the electric field coupling strength $d_{\perp}$.

Under an electric field $E_\zeta$, we expect NV phase accumulation $\phi_{NV} = 2\pi d_{\perp}  E_\zeta  \tau$ (the initial $2\pi$ is to convert the angle unit to rad). After a $\frac{\pi}{2}$ pulse, this phase signal is converted to the probability of being in the $|0\rangle$ state, denoted by $P$. With an initial $( \frac{\pi}{2} )_x$ pulse and a $( \frac{\pi}{2})_y$ pulse at the end,    
\begin{eqnarray}
P = \frac{1}{2} ( 1-  \sin{ (\phi_{NV})} ) = \frac{1}{2} ( 1-  \sin{ (2\pi d_{\perp}   E_\zeta  \tau)}),
\end{eqnarray}
Suppose a small electric field $\delta E$ along $\hat{\zeta}$, the total number of photons collected due to $\delta E$ after $N_{avg}$ number of averages is: 
\begin{eqnarray}
\textrm{signal} = \left( \frac{\partial P}{\partial E}\right) _{E=0} \cdot \delta E \cdot C \cdot F T_r \cdot N_{avg} =  \pi d_{\perp} \tau \cdot \delta E \cdot C \cdot F T_r \cdot N_{avg}, 
\end{eqnarray}
The photon shot noise is 
\begin{eqnarray}
\textrm{noise}=\sqrt{FT_r \cdot N_{avg}}
\end{eqnarray} 
Within 1 second of integration time, $N_{avg} = 1/(t_{ini}+\tau)$. Thus we obtain the signal-to-noise ratio (SNR) within 1 sec integration time:
\begin{eqnarray}
\textrm{SNR} =  \pi d_{\perp} \tau \cdot \delta E \cdot C \cdot \sqrt{ \frac{F T_r}{t_{ini}+\tau}},
\end{eqnarray} 
The sensitivity $\eta_{E}$ is defined as the minimal $\delta E$ that can be detected with SNR = 1 within 1 sec, hence
\begin{eqnarray}
\eta_{E} = \sqrt{ \frac{t_{ini}+\tau}{F T_r}} \frac{1}{\pi d_{\perp} \tau C}
\end{eqnarray} 
In AC electric field imaging, we have $F$ = 100 kcounts/sec,  $\tau$ = \SI{8}{\mu s}, $C \approx 0.2$, $T_r$ = 200 ns, and  $t_{ini}$=\SI{2}{\mu s}. Plugging these experimental parameters and $d_\perp=$ \SI{0.17}{\,MHz/V/\mu m} into Equation (5), we obtain: $\eta_{E}$ = \SI{26}{\,mV/\mu m/\sqrt{Hz}} = \SI{260}{\,V/cm/\sqrt{Hz}}, which is comparable to the sensitivity obtained by a single NV in bulk diamonds \cite{dolde2011electric}. 

In our motion-enabled DC electric field imaging, the directly measured signal is dominated by the local spatial gradient, so we examine the electric field gradient sensitivity: $\eta_{gradient} = \frac{\eta_{E}}{A}$, where $A$ is the tip oscillation amplitude. In our experiments, $A=13$ nm, therefore $\eta_{gradient}$ = \SI{2000}{\,mV/\mu m^2/\sqrt{Hz}} = \SI{2}{\,V/\mu m^2/\sqrt{Hz}}.

\subsection{E. Multiple-pillar diamond probe and scanning details}

\begin{figure}[htbp!]
\centering
\includegraphics[scale=0.51]{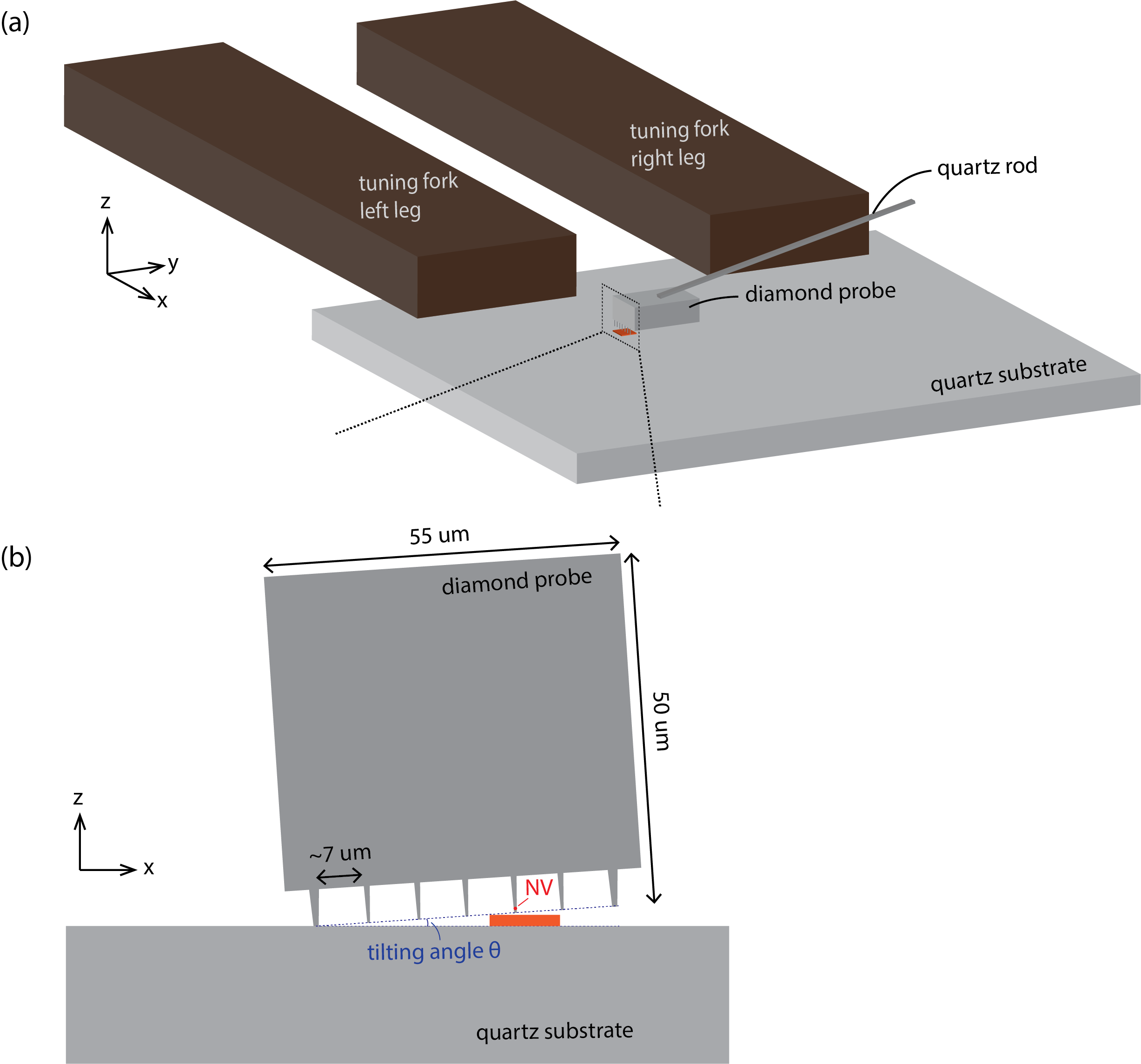}
 \caption{\label{fig:multipillar_probe} (a) The probe is glued to the right leg of the tuning using a quartz rod. (b) Side view of the probe with multiple pillars. The outer two pillars are $\sim$800 nm in diameter. One of them is in direct contact with the substrate and performs AFM scanning. The middle five pillars are 300 nm in diameter. One of them is close to the sample of interest and performs NV quantum sensing. The NV-sample distance is hence controlled by the tilting angle $\theta$, which is calibrated carefully. }
 \end{figure}

\begin{figure}[htbp!]
 \centering
\includegraphics[scale=0.5]{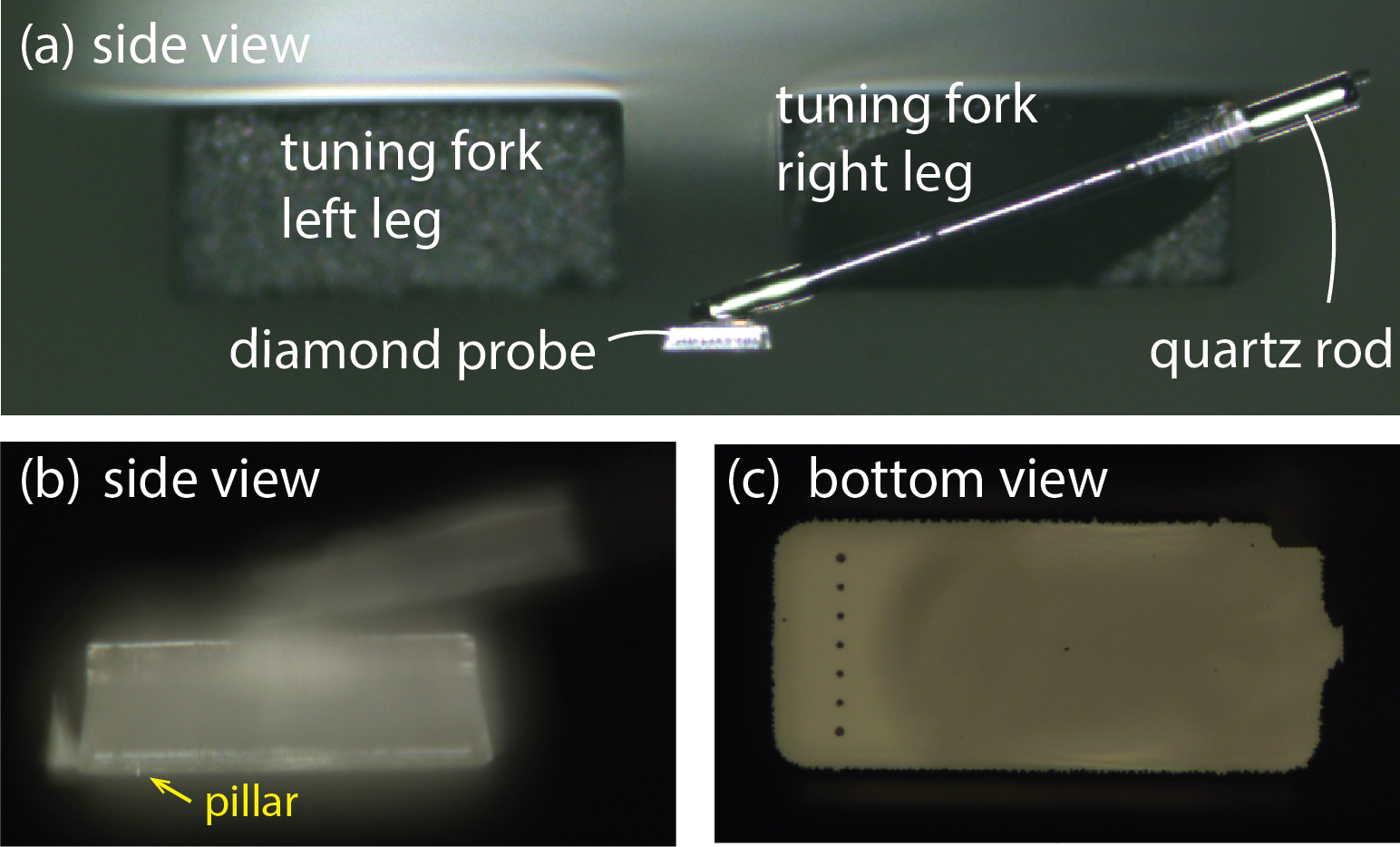}
 \caption{\label{fig:probe_pics} Optical microscope images of the tuning fork and the diamond probe. }
 \end{figure}

Fabrication of the diamond probe follows the procedure descried in \cite{zhou2017scanning}. To improve the probability of finding good single NVs in a probe and scanning stability, we fabricated multiple pillars on a single probe as shown in Figure~\ref{fig:multipillar_probe} and Figure~\ref{fig:probe_pics}. There are 7 pillars in a row. The outer two pillars have a diameter of $\sim$800 nm, and the middle five ones are 300 nm in diameter. During the scanning experiments, the probe is slightly tilted such that one of the outer thick pillar is in direct contact with the substrate which performs AFM scanning, and one of the middle thin pillar hovers above the sample of interest at a distance of 20-40 nm which performs the sensing task. The tilting angle $\theta_B$ is controlled by a manual goniometer (Edmund). This method also prevents the sensing pillar from picking up dirt or scratching the sample, and the thick AFM scanning pillar is less likely to break during harsh long-hour scanning. Overall, this multi-pillar scanning method improves the lifetime of a diamond probe. 

The tilting angle $\theta$ is carefully calibrated in advance. We start with a large $\theta$ and perform a 1D scan over two metal gaps to measure the signal profile. Based on the spatial resolution and signal strength, we gradually decrease $\theta$, and repeats the 1D scan along the same line. As shown in Figure 3c of the main text, this process is then repeated multiple times until the signal profile is consistent with simulated signal at a distance of $<$100 nm away by COMSOL. More details on COMSOL simulation is included in section III. 

\subsection{F. Motion-enabled DC imaging}

\begin{figure}[htbp!]
\centering
\includegraphics[scale=0.5]{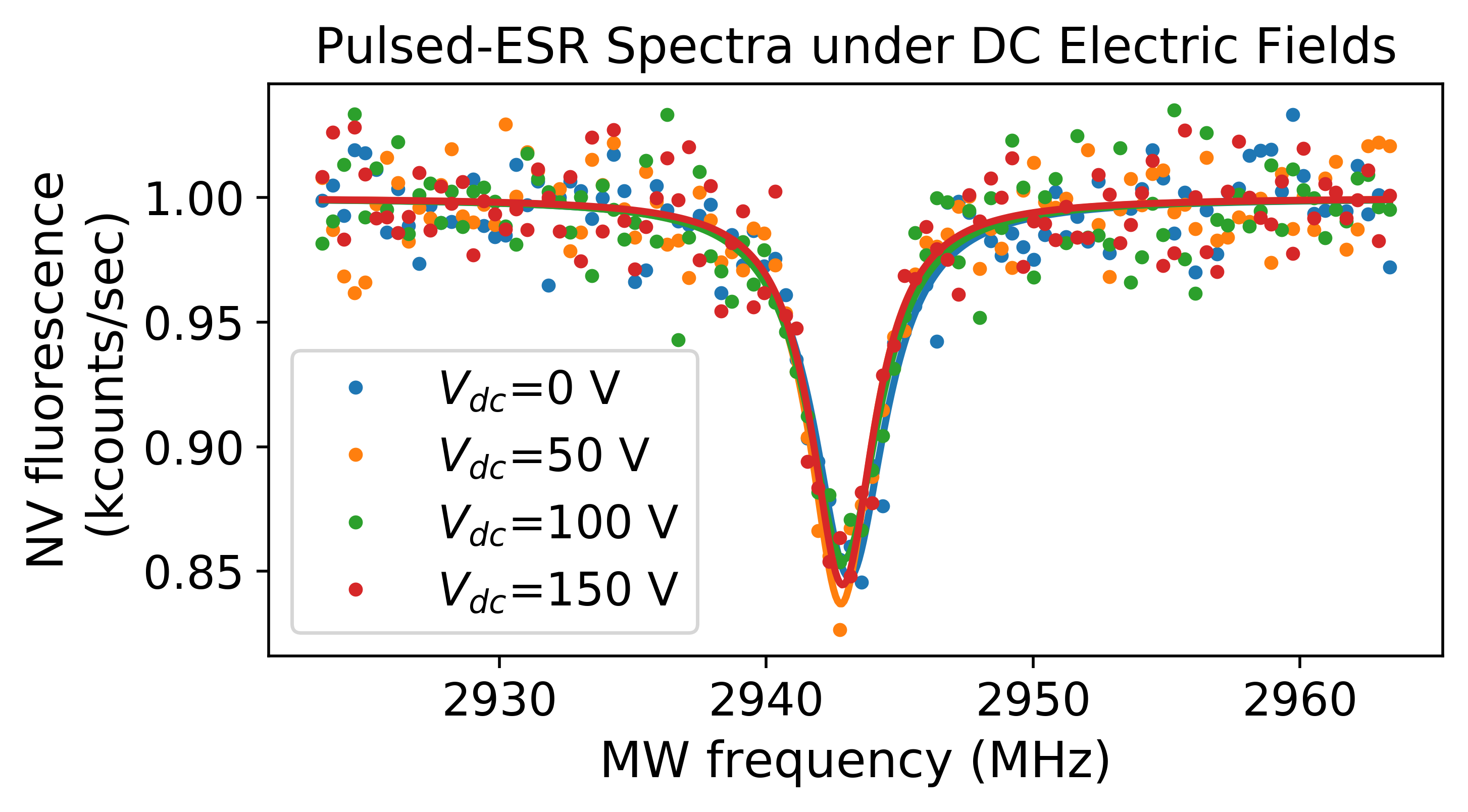}
\caption{\label{fig:DC_sensing} Pulsed-ESR spectra showing no discernible shifts upon applying DC voltages}
\end{figure}

As mentioned in the main text, we observed a strong electric screening effect at DC and low frequencies. Figure~\ref{fig:DC_sensing} shows the pulsed-ESR spectra measured in the presence of DC voltages applied across the 500 nm gap. No discernible shifts were observed. 

\begin{figure}[htbp!]
\centering
\includegraphics[scale=0.36]{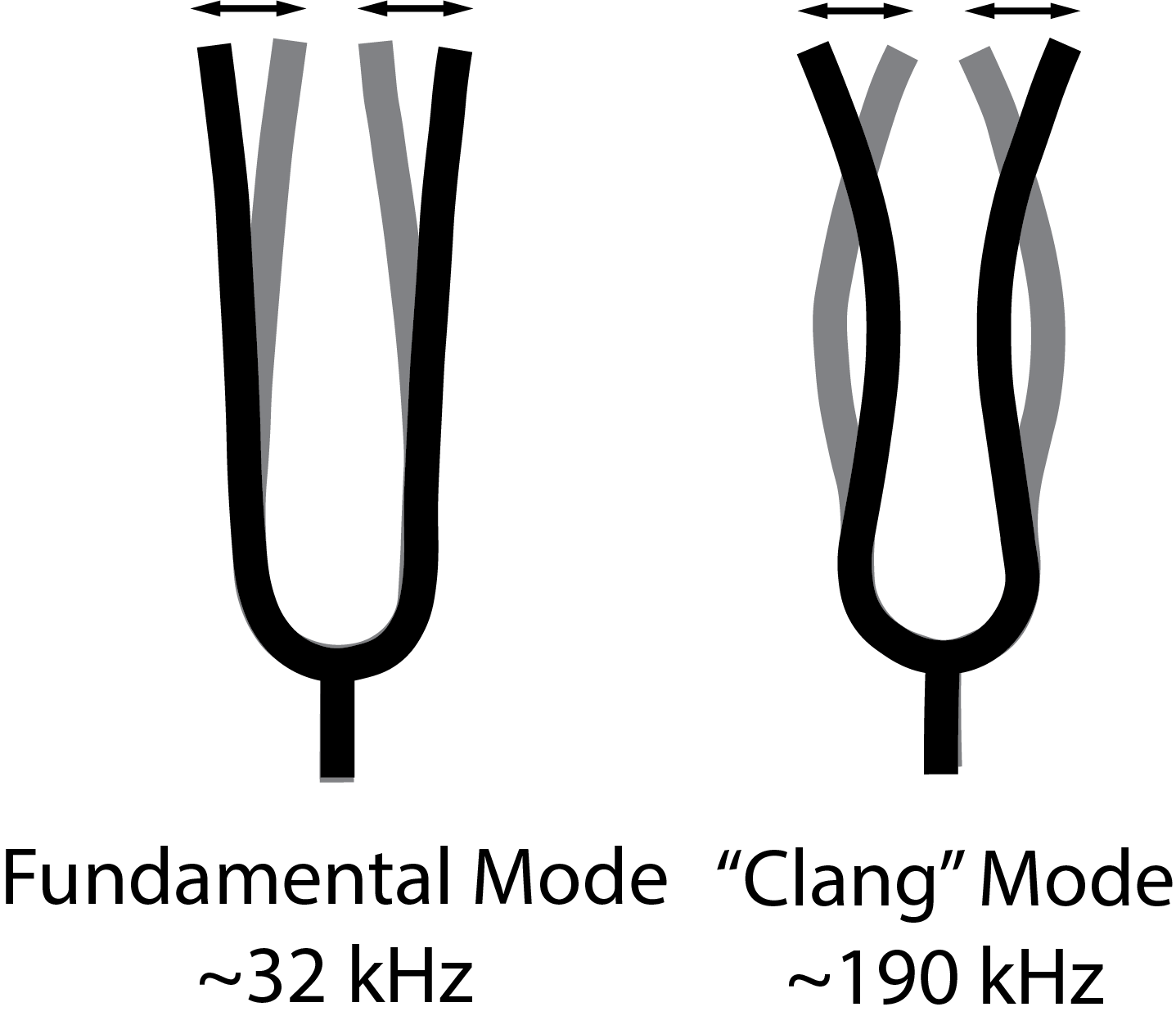}
\caption{\label{fig:tuning_fork} Two modes of a tuning fork. The fundamental mode is used in AC imaging, and the 'clang' mode is used in DC imaging.}
\end{figure}

To overcome the electric screening effect at low frequencies, we mechanically oscillate our diamond probe in imaging DC electric field signals. The probe is attached to one of the legs of a quartz tuning fork (Figure~\ref{fig:multipillar_probe} and Figure~\ref{fig:probe_pics}), and a resonant electric signal excites the motion of the tuning fork (Digi-Key 1123-ND) due to the piezoelectric properties of quartz. The fundamental mode of the tuning fork is about 32 kHz, and its clang mode is about 190 kHz, roughly 6 times higher than the fundamental (Figure~\ref{fig:tuning_fork}). As shown in Figure 2 of the main text, the electric screening effect is still severe at 32 kHz, but significantly reduced at about 190 kHz, hence we excite the tuning fork at its clang mode. 

The probe motion increases as the excitation $V_{pp}$. However, the quantitative relationship varies between different tuning forks and needs to be calibrated every time when we switch a probe.  In our experiments, the maximum probe motion amplitude is $<$20 nm, as estimated by comparing the detected signal distribution with COMSOL simulation.

\section{III. COMSOL simulation} 
The finite-element calculation package COMSOL performs the electrostatic simulations.

\subsection{A. Dielectric screening}
\begin{figure}[htbp!]
\centering
\includegraphics[scale=0.48]{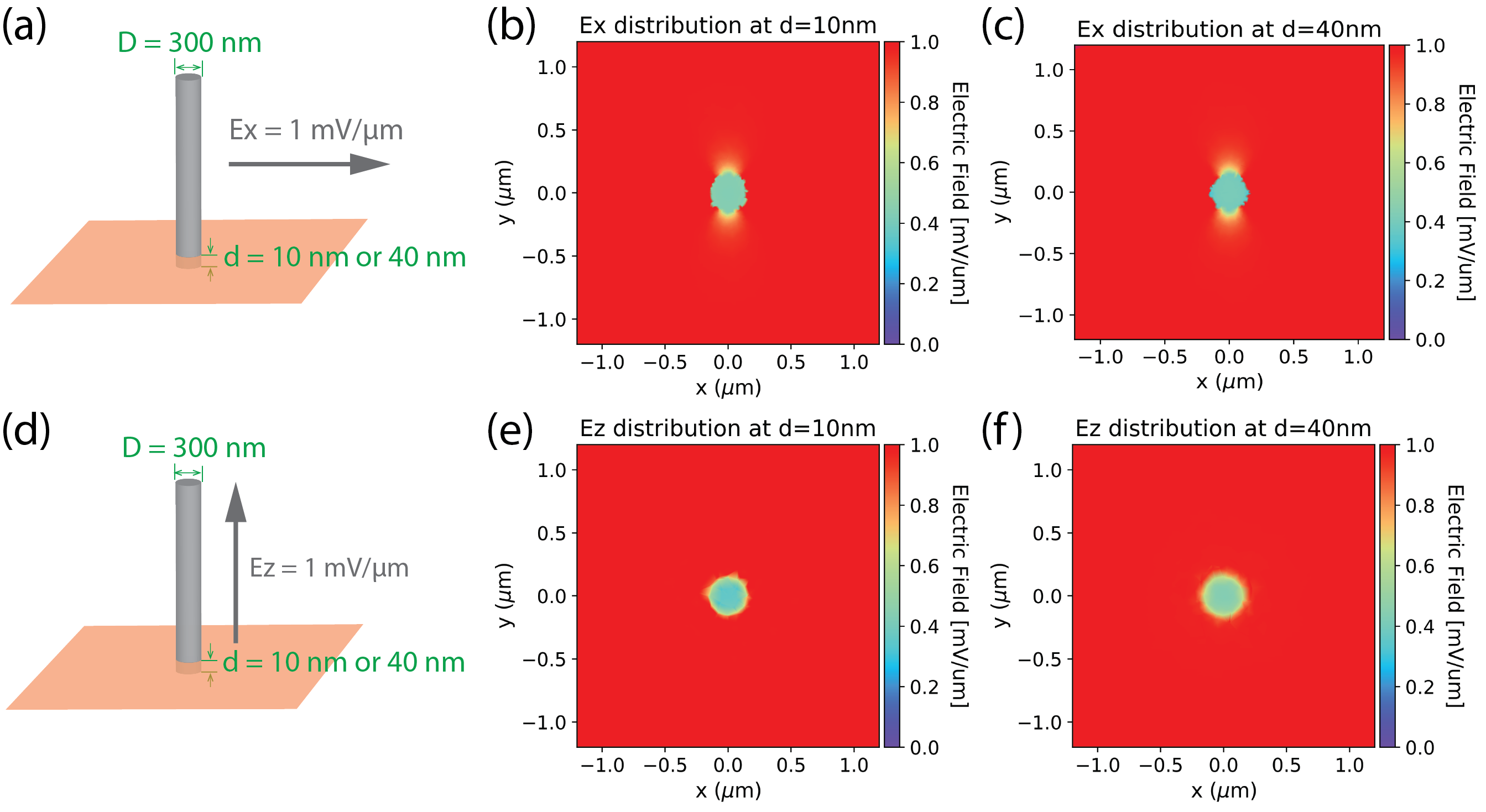}
\caption{\label{fig:dielectric_screening} COMSOL simulation of the dielectric screening effect in a diamond tip, with a flat top surface and 300 nm in diameter. (a) The probe in the presence of an external horizontal electric field of 1 $mV/\mu m$. (b) and (c) plot the 2D distribution of the simulated electric fields at depths of 10 nm and 40 nm from the tip surface respectively. (d) The external field is vertical with a magnitude of 1 $mV/\mu m$. (e) and (f) plots the simulated electric fields at depths  of 10 nm and 40 nm respectively.  }
\end{figure}

First, we considered dielectric screening due to the non-negligible relative permittivity of diamond, which is $\sim$5.7. Our diamond scanning tip has a circular flat top with a diameter of 300 nm, and the NV is located at $\sim$40 nm away from the surface. Figure~\ref{fig:dielectric_screening} visualizes the simulated electric fields inside and outside the diamond. $E_{x,y}$ and $E_z$ at the location of NV are both reduced to $\sim$41$\%$ as compared to the field magnitudes in the air. In subsequent discussions, we take into account this reduction to compare between the simulation and experimental data without explicitly mentioning this.

\subsection{B. Simulation of AC and DC electric field distributions}
\begin{figure}[htbp!]
\centering
\includegraphics[scale=0.6]{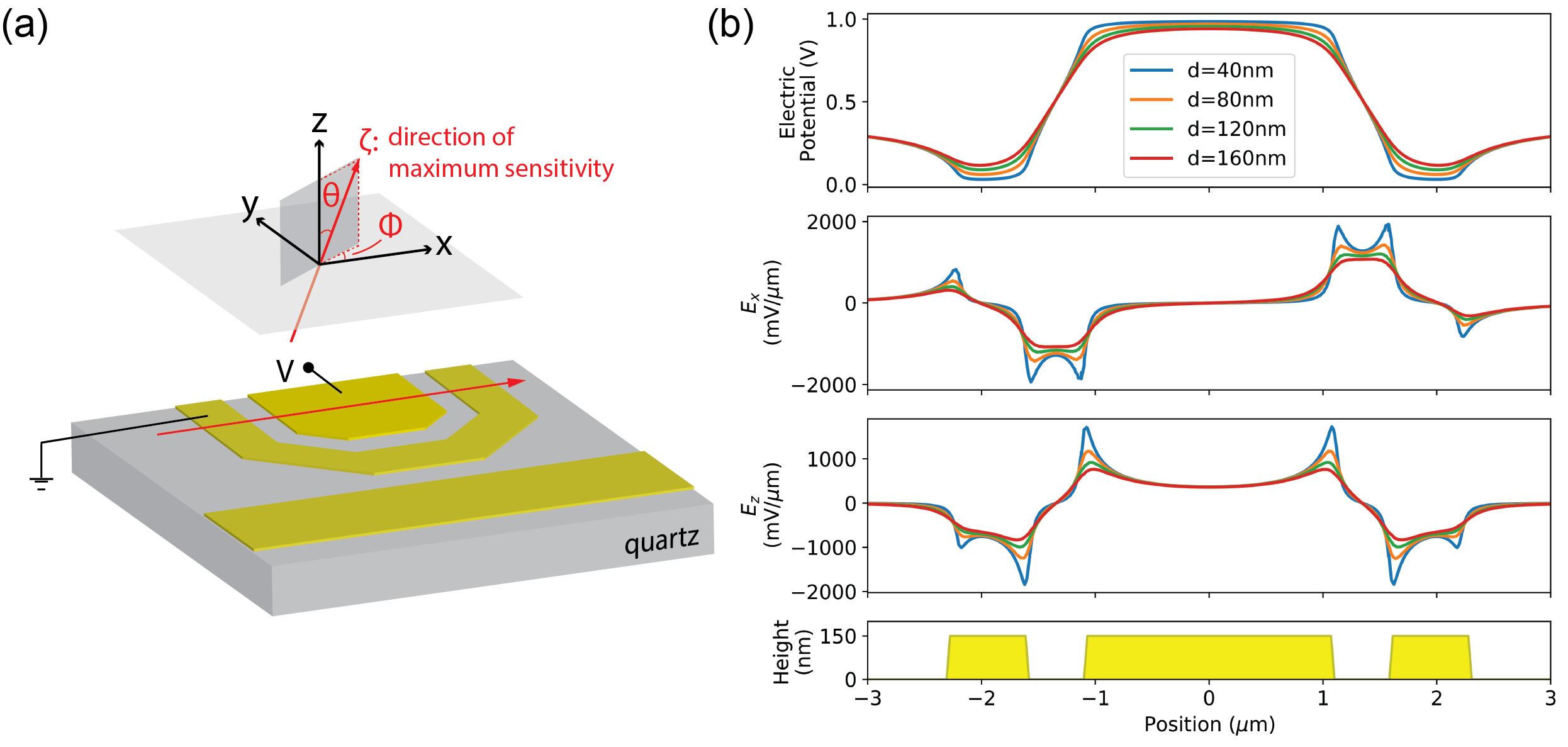} 
\caption{\label{fig:COMSOL_schematic} COMSOL simulation of the electric potential and electric field profiles along a 1D line, represented by a red arrow in (a). The first three plots in (b) show the simulated profiles at several distances, and the fourth plot shows the metal height profile. The actual electric field detected by the NV is the component along $\zeta$, which is a linear combination of $E_x$ and $E_z$. }
\end{figure}

To simulate our device, we used the real device dimension to model the geometry. In 3D simulations, The 2D lithography design was imported and extruded for a thickness of 150 nm to create 3D structures. The NV electric field sensitivity is maximized along a particular direction, which depends on the NV orientation and bias magnetic field direction. As shown in Figure 1 of the main text, the energy shift caused by electric field is $ \propto E_\perp \cos{(2\phi_B+\phi_E)}$, so the direction of maximum sensitivity is at $\phi_E=-2\phi_B$, represented by $\hat{\zeta}$. In COMSOL simulation, we first calculated the 2D distribution of all the three components of electric fields $E_x$, $E_y$ and $E_z$ at different distances $h$ from the sample surface in the air, then calculated the $\hat{\zeta}$ component: $E_{\hat{\zeta}} = (E_x \cos{\phi} + E_y \sin{\phi}) \sin{\theta} + E_z \cos{\theta}$, where $\phi$ is the azimuth angle and $\theta$ is the zenith angle as depicted in Figure~\ref{fig:COMSOL_schematic}. 

In imaging AC electric fields, the signal is modulated at $\sim$250 kHz. In imaging DC electric fields, both the probe and signal oscillate at $\sim$ 190 kHz. $\phi$, $\theta$ and the distance $h$ are the three free parameters to produce Figure 3e of the main text, where we used $\phi=20\degree, \theta=45\degree, h=$ 90 nm. In DC imaging, a XY4-based AC sensing pulse sequence is synchronized with the motion of the diamond probe. Therefore, the coherent phase accumulation is proportional to the product of the electric field gradient along the probe oscillation direction and the oscillation amplitude. These are two additional free parameters in producing Figure 4e of the main text, the oscillation direction is set to be at $30\degree$ relative to the scanning direction  $\hat{x}$  and the oscillation amplitude is 13 nm.
\nocite{*}
\bibliography{si_reference}